\documentclass[12pt]{article}

\usepackage[english]{babel}
\usepackage[utf8]{inputenc}
\usepackage{lmodern,csquotes}
\usepackage[T1]{fontenc}

\usepackage[a4paper,left=0.9in,right=0.9in,top=1.3in,bottom=0.8in]{geometry}

\usepackage[font=small,labelfont=bf]{caption}
\usepackage[style=numeric, sorting=nyt, maxcitenames=1,maxbibnames=99]{biblatex}
\usepackage{amssymb, amsmath, amsthm, mathtools, dsfont, xfrac, bm}
\usepackage{float}
\usepackage{eurosym}
\usepackage{graphicx, tikz, pgfplots, pgfplotstable, booktabs, multirow, enumitem}
\pgfplotsset{compat=newest}
\usepgfplotslibrary{fillbetween}

\usepackage[section]{placeins}

\definecolor{mylightyellow}{rgb}{1,1,.8}
\definecolor{mylightgreen}{rgb}{.8,1,.8}
\definecolor{mydarkorange}{RGB}{217,126,29}
\definecolor{mydarkred}{RGB}{178,34,34}
\definecolor{mydarkgreen}{RGB}{34,139,34}
\definecolor{mydarkblue}{RGB}{72,61,139}
\definecolor{mydarkyellow}{RGB}{218,165,32}

\usepackage[normalem]{ulem}
\newcommand\soutpars[1]{\let\helpcmd\sout\parhelp#1\par\relax\relax}
\long\def\parhelp#1\par#2\relax{%
  \helpcmd{#1}\ifx\relax#2\else\par\parhelp#2\relax\fi%
}

\usepackage{hyperref}
\hypersetup{
 colorlinks=true,breaklinks=true,
 urlcolor= mydarkblue,linkcolor= mydarkblue,citecolor= mydarkgreen,
 pdfauthor={Corti, S., Daluiso, R., Pallavicini, A.}
}

\theoremstyle{plain}

\theoremstyle{definition}
\newtheorem{definition}{Definition}
\theoremstyle{remark}

\DeclarePairedDelimiterX\QCond[2]{\{}{\}}{\,#1\,\delimsize\vert\,\mathopen{}#2\,}
\DeclarePairedDelimiterX\QUncond[1]{\{}{\}}{\,#1\,}
\DeclarePairedDelimiterX\ECond[2]{[}{]}{\,#1\,\delimsize\vert\,\mathopen{}#2\,}
\DeclarePairedDelimiterX\EUncond[1]{[}{]}{\,#1\,}

\begin{document}

\title{Optimal strategy and deep hedging \\ for share repurchase programs\footnote{The authors report no potential competing interests. The opinions expressed in this document are solely those of the authors and do not represent in any way those of their present and past employers.}}

\author{
S.~Corti\thanks{Politecnico di Milano, Department of Mathematics. Address: Piazza Leonardo da Vinci 32, Milano 20133, Italy. Email address: \texttt{stefano7.corti@mail.polimi.it}.},
R.~Daluiso\thanks{Intesa Sanpaolo, Financial Engineering. Address: largo Mattioli 3, Milano 20121, Italy. Email address: \texttt{roberto.daluiso@intesasanpaolo.com}.},
A.~Pallavicini\thanks{Intesa Sanpaolo, Financial Engineering. Address: largo Mattioli 3, Milano 20121, Italy. Email address: \texttt{andrea.pallavicini@intesasanpaolo.com}.}}

\date{
\small First Version: December 9, 2025.  This version: \today
}

\maketitle
\thispagestyle{empty} 

\begin{abstract}
In recent decades, companies have frequently adopted share repurchase programs to return capital to shareholders or for other strategic purposes, instructing investment banks to rapidly buy back shares on their behalf. When the executing institution is allowed to hedge its exposure, it encounters several challenges due to the intrinsic features of the product. Moreover, contractual clauses or market regulations on trading activity may make it infeasible to rely on Greeks. In this work, we address the hedging of these products by developing a machine-learning framework that determines the optimal execution of the buyback while explicitly accounting for the bank’s actual trading capabilities. This unified treatment of execution and hedging yields substantial performance improvements, resulting in an optimized policy that provides a feasible and realistic hedging approach. The pricing of these programs can be framed in terms of the discount that banks offer to the client on the price at which the shares are delivered. Since, in our framework, risk measures serve as objective functions, we exploit the concept of indifference pricing to compute this discount, thereby capturing the actual execution performance.
\end{abstract}

\bigskip

\noindent {\bf AMS classification codes:} 65C05, 91G20, 91G60.

\newpage

\pagestyle{myheadings}
\markboth{}{%
  \footnotesize
  \makebox[0pt][l]{\parbox{0.8\textwidth}{%
    Corti, Daluiso, Pallavicini\\
    Optimal strategy and deep hedging for share repurchase programs
  }}%
}

\section{Introduction}
\label{sec:intro}
Payout policies play a significant role in corporate finance literature. Generally, to return capital to their shareholders, companies can use either dividends or share repurchases. Modigliani--Miller \cite{Miller3} theorem states that the two approaches are equivalent in the absence of taxes, market friction, and information asymmetry. However, in practical settings \cite{Allen4, Farre-Mensa5}, share repurchases are sometimes preferred for strategic reasons, such as tax considerations or to adjust the firm’s capital structure. Furthermore, these programs can signal stock undervaluation and deter takeovers.

\vspace{0.1 cm}
Briefly, under these contracts, the company agrees with an investment bank on a predetermined volume of shares to buy within a specified maturity date and on a benchmark price. The bank is responsible for the market operations and earns the difference between the benchmark price, typically an average of prices measured on the repurchase days, often offered at a discount, and the average execution price. The program is usually divided into two phases, the second of which begins when the financial institution is granted the right to suspend trading, leading to the final settlement. The speed with which repurchases are carried out has earned them the name Accelerated Share Repurchases, or ASR.

\vspace{0.1 cm}
From an operational point of view, the bank determines daily how many shares to repurchase and when to exercise the termination right. In the literature \cite{Gueant6, Gueant12, Gueant13}, the entire process is framed as a control problem, since profit maximization requires the executor to behave according to an optimal policy. The framework is characterized by a high-dimensional state space, as the path-dependent payoff makes day-by-day decisions dependent on the underlying asset value and on the variables that determine the final reward, namely the cumulative quantity of shares traded, the total expenditure, and the price process of the benchmark.

\vspace{0.1 cm}
From the bank's perspective, the dependence of the cash flow at the termination date on the equity price creates an exposure to the underlying asset. In some ASR contracts, the financial institution is allowed to hedge its position, a task that proves highly complex due to the barrier-like features embedded in the product. For instance, the payoff structure may suggest a strategy based on sudden switches between extreme repurchase regimes: it may be optimal to buy larger quantities on the market when the value of the underlying falls below its average price while buying the minimum admissible quantity in the opposite scenario. These discontinuities, sometimes also caused by the policy that defines contract termination, make ASR a barrier-type product. Consequently, its Greeks show highly volatile dynamics, with frequent changes in sign and large values in absolute terms, and their computation may be affected by numerical instability. Relying on such highly variable signals can lead to abrupt adjustments in the hedging portfolio, resulting in significant transaction costs due to bid–ask spreads and substantial replication errors when rebalancing is performed at discrete dates.

\vspace{0.1 cm}
Furthermore, trading constraints introduce additional limitations to the use of Greeks. While some clauses apply solely to the buyback activity, others simultaneously affect both the units of the underlying asset delivered to the client and those used for risk management, as they limit the total number of shares that can be purchased daily to a percentage of the market volume. As a result, the position suggested by the Greeks, computed ex post, may be infeasible whenever total trading exceeds this threshold. Therefore, the optimal policy should incorporate the agent’s effective operational capacity, which requires embedding the hedging activity directly into the control problem and thereby making it interdependent with the execution of the program.

\vspace{0.1 cm}
This work focuses on the full management of share repurchase programs. In particular, we investigate whether a unified framework for handling ASR contracts and their associated hedging can yield meaningful performance improvements in real terms. Moreover, modeling an agent that jointly determines how to manage the contract and how to hedge the resulting position enables the implementation and optimization of policies that explicitly account for actual market constraints. By incorporating trading capacity directly into the control problem, the resulting strategy offers a more feasible and realistic approach to managing the bank’s exposure.

\vspace{0.1 cm}
We adopt machine learning techniques, which are naturally suited to handling high-dimensional problems and numerous constraints that often challenge traditional methods. Following an approach similar to \cite{Baldacci10}, we define several classes of strategies characterized by a set of parameters, so that determining the optimal behavior reduces to searching for the optimal parameter vector. Specifically, we delegate both the design of repurchase and hedging policies to neural networks, leveraging their computational capabilities and flexibility, which allows us to feed the state-space variables directly into the control function. In this direction, we build on the work of O. Guéant, I. Manziuk, and J. Pu \cite{Gueant6}, who apply neural networks to ASR programs, and on that of H. Buehler, L. Gonon, J. Teichmann, and B. Wood \cite{Buehler7}, who exploit them to hedge derivative portfolios.

\vspace{0.1 cm}
In a machine learning framework, the optimization procedure relies on the specification of a loss function. In our setting, this role is played by risk measures \cite{Foellmer14}, as they provide a natural way to assess the performance of policies. Furthermore, they allow us to incorporate the concept of \textit{indifference pricing} \cite{Buehler7}, offering a meaningful alternative to the price obtained using the traditional risk-neutral valuation formula. The latter loses relevance in this context, as it relies on the ability to construct and follow a replicating portfolio defined through the Greeks, an approach that, as previously discussed, is often infeasible due to operational constraints or high transaction costs.

\vspace{0.1 cm} 
In our analysis, we require the payoff to depend differentiably on the networks’ outputs in order to leverage classical back-propagation algorithms. We adopt the same approach as in \cite{Gueant6}, where the authors introduce a relaxation of the optimal stopping problem by replacing the concept of optimal stopping time with a probabilistic representation. This transforms the exercise decision from discrete to continuous, enabling gradient-based optimization.

\vspace{0.1 cm}
From a mathematical perspective, share repurchase programs have been already examined in the literature. In \cite{Jaimungal32}, the authors analyze an ASR contract with a fixed number of shares in a continuous-time model with a quadratic running penalty on remaining shares. The optimal execution policy is reduced to a partial differential equation, which is then solved numerically. Moreover, by considering the ratio between the underlying price and its running average since the inception of the contract, they succeed in reducing the dimensionality of the equation from five to three. A different approach is taken in \cite{Gueant12}, where the optimal execution problem is studied in discrete time by a risk-averse agent within an expected-utility framework. Starting from the Bellman equation, the authors develop a pentanomial tree to determine the optimal strategy. A similar methodology is employed in \cite{Gueant13} for fixed-notional ASR contracts. Finally, as mentioned above, neural network based methods have also been explored for managing these products in \cite{Gueant6}.

\vspace{0.1 cm}
The hedging technique based on deep neural networks presented in \cite{Buehler7}, and usually referred to as "deep hedging", provides the complete theoretical framework for applying machine learning, specifically neural networks, to the hedging problem, relying on the concept of convex risk measures. This approach has been further developed in many directions. For instance, it has been extended beyond traditional delta hedging to manage higher-order risks and implied volatility dynamics. Some examples are \cite{Horikawa22}, comparing deep hedging with risk-neutral delta hedging, \cite{Francois23}, which applies deep hedging using the implied volatility surface for gamma hedging, \cite{Armstrong24}, focusing specifically on deep gamma hedging, and \cite{Cao25}, which employs deep distributional reinforcement learning to hedge both gamma and vega effectively.

\vspace{0.1 cm}
To our knowledge, deep hedging has only been applied to the hedging of traditional derivative portfolios, obtaining satisfactory results; for example, in \cite{Mikkila8} the authors deal with options using intra-day data from actual markets, while \cite{Oya17} deals with Bermudan swaptions and \cite{Pickard16} with American put options. We are motivated by the question of whether this technique can be successfully applied to a more complex derivative-like product. If this proves to be the case, then it is straightforward to unify the repurchase strategy and its hedging into a joint optimization framework. As explained above, this would enable a complete workflow to determine optimal behavior, including both market risk protection and contractual constraints. 

\vspace{0.1 cm}
We conclude by outlining the structure of the article. Section 2 presents the mathematical setting, the contract features, clauses, and the payoff, followed by the formulation and relaxation of the control problem and a discussion of how indifference pricing applies to ASR. Section 3 addresses the management of repurchase programs, implementing policies under the relaxed framework that enables gradient-based optimization. Section 4 then applies the hedging framework to these policies, evaluating the effectiveness of the overall approach.

\section{The model}
\label{sec:model}
\subsection{Mathematical Setting}
In our analysis, in order to simplify the exposition, we assume that buyback operations take place on a discrete set of equally-spaced dates. This assumption induces a discrete time grid with step size $\delta t$, which we set to one day. Each point in the grid $t_n = n \cdot \delta t$  corresponds to the $n\text{-th}$ day. The program starts at $t_1 = \delta t$, while the maturity is denoted by $t_{Max} = N_{Max} \cdot \delta t$. The earliest settlement date is $t_{Min} = N_{Min} \cdot \delta t$ and it sets the beginning of the exercise window.

\vspace{0.1 cm}
To simplify the notation, we define the index sets $\mathcal{T}= \{ n \in \{1, ...,N_{Max}\} \}$, representing the full duration of the program, and $\mathcal{E}= \{ n \in \{N_{Min}, ...,N_{Max}\} \}$, corresponding to the exercise window.

\vspace{0.1 cm}
We proceed to define the probability space $(\Omega, \mathbb{P}, \mathcal{F})$, where $\mathbb{P}$ is the physical measure. The underlying asset is modeled by the stochastic process $(S_n)_{n \in \mathcal{T}}$, adapted to the filtration $(\mathcal{F}_n)_{n \in \mathcal{T}}$. 

\vspace{0.1 cm}
In our numerical experiment, we adopt the Black–Scholes model for the underlying price process $(S_n)_{n \in \mathcal{T}}$, as our focus is on the design of policies rather than on the specific dynamics of $S_t$. This choice also enables efficient sampling from $\Omega$ to construct mutually independent training, validation, and test sets. Following standard practice in the literature \cite{Baldacci10}, we assume that the physical measure $\mathbb{P}$ coincides with the risk-neutral measure $\mathbb{Q}$. Furthermore, since buyback programs typically last only a few months, interest rates have a negligible effect; therefore, we set the risk-free rate $r$ to zero.

\vspace{0.1 cm}
The bank’s repurchases are described by the following two $\mathcal{F}_n$-adapted processes and by a $\mathcal{F}_n$-stopping time $\tau = n^* \cdot dt$, with $n^* \in \mathcal{E}$:
\begin{itemize}[noitemsep, topsep=0pt]
\item $(b_n)_{n \in \mathcal{T}}$, the number of shares bought each day for the buyback program.
\item $(h_n)_{n \in \mathcal{T} \setminus \{ N_{Max} \}}$, the equity units held at each point along the time grid for hedging purposes.
\end{itemize}

\vspace{0.1 cm}
For each $n \in \mathcal{T}$, let $Q_n$ and $W_n$ denote, respectively, the cumulative number of repurchased shares and the corresponding cash outflow up to and including day $n$. We also define the benchmark price process $(A_n)_{n \in \mathcal{T}}$, given by the average price starting from $t_1$. The dynamics of these processes are given by:
\begin{equation*}
Q_n = \sum_{k = 1}^{n} b_k, \qquad W_n = \sum_{k = 1}^{n} b_k \cdot S_k \quad \text{ and } \quad A_n = \frac{1}{n} \cdot \sum_{k = 1}^{n} S_k.
\end{equation*}

\vspace{0.1 cm}
In these setting, we are ignoring intraday price movements and transaction costs. In particular, we assume that at the time of the repurchase decision, the price at which the transaction is executed is known exactly and coincides with the price used in the arithmetic average defining the payoff. Ignoring intraday variations is expected to have a negligible impact, as they occur on a much shorter time scale than the duration of the product, while transaction costs can be easily incorporated into the model.

\subsection{Term Sheet}
Before presenting the analytical expression of the Profit and Loss, we first outline the term sheet of the buyback under analysis. We assume that the volume of shares is expressed as the total amount of cash to be spent on the market, though one could repeat the analysis for contracts defining it as the number of shares to repurchase.

\vspace{0.1 cm}
We also include one of the most common features in ASR. During the second phase, the bank has the right to exercise when the notional amount falls within a predefined range $[W_{Min}, \text{ } W_{Max}]$. Its width, known as the \textit{greenshoe}, is one of the most used tools to make ASR more competitive, for example, by offering a larger discount on the benchmark price. 

\vspace{0.1 cm}
We now briefly consider the timing of share delivery and payments. It is common for the company to gradually provide the necessary funds over the life of the program, while the bank delivers the shares as they are purchased. The final payoff is made at the termination date of the contract, denoted by $\tau$, hence the term post-paid ASR. Alternative delivery mechanisms should have little impact on the pricing of the contract given its comparably low maturity, and would anyway be completely irrelevant under our assumption of zero interest rates.

\vspace{0.1 cm}
Finally, we address the trading constraints, which play a major role in our model. Specifically, daily trading is typically limited to a percentage of the volume of the asset exchanged on the market. The processes $(\underline{\nu}_n)_{n \in  \mathcal{T}} \text{ and } (\overline{\nu}_n)_{n \in  \mathcal{T}}$ denote the lower and upper bounds for the buyback portfolio, and their computation usually excludes end-of-day auctions from the total market volume. Instead, auctions are incorporated into $(\overline{\nu}_n^h)_{n \in  \mathcal{T} \setminus \{ N_{Max} \}}$, which also limits the hedging portfolio and may prevent the implementation of the Greeks. The first resulting limitation is imposed by the ASR contract, while the second arises from market regulations:
\begin{equation}
\label{eq:constraints}
\left\{
\begin{aligned}
  &b_n \in [\,\underline{\nu}_n,\, \overline{\nu}_n]\\[0.5em]
  &b_n + |h_n - h_{n-1}| \leq \overline{\nu}_n^h
\end{aligned}
\right.
\qquad \forall n \in \mathcal{T}.
\end{equation}

\subsection{Profit and Loss}
We now introduce the bank's profit and loss, or $PnL$, which consists of two components: the payoff of the ASR and, when allowed, the result of the hedging activity. The corresponding hedging portfolio is automatically closed once the repurchase program is completed.

\vspace{0.1 cm}
Both cash flows occur at the final settlement $\tau = n^* \cdot \delta t$. The total profit and loss is therefore
\begin{equation}
PnL_{\tau} = PnL^{ASR}_{\tau} + PnL^{Hedge}_{\tau}.
\end{equation}

\vspace{0.1 cm}
We now detail the analytical expression of $PnL^{ASR}_{\tau}$.
\begin{equation}
\left\{
\begin{alignedat}{2}
  &PnL^{ASR}_{\tau} = (1 - \delta) \cdot A_{n^*} \cdot Q_{n^*} - W_{n^*} \\
  &W_{n^*} \in [W_{Min}, W_{Max}]
\end{alignedat}
\right.
\end{equation}
The parameter $\delta$ denotes the discount offered to the client on the benchmark $A_{n^*}$, whose discounted value $(1 - \delta) \cdot A_{n^*}$ can be interpreted as the per-share delivery price and determines the bank's remuneration. The profit depends on the ability to execute the repurchases $b_n$ at favorable prices.

\vspace{0.1 cm}
To characterize the dynamics of the problem, we define the state space at time $t_n$ as the collection of variables entering the definition of $PnL^{ASR}$ and containing all the information relevant for managing the contract:
\begin{equation*}
\mathcal{S}_n = (n, \text{ }S_n, \text{ }A_n, \text{ } Q_{n-1}, \text{ }W_{n-1}).
\end{equation*}

\vspace{0.1 cm}
We observe that, in theory, it is not possible to guarantee that $ \exists n \in \mathcal{E}: \ W_{n} \in [W_{Min}, W_{Max}]$ with probability $\mathbb{P} = 1$. Indeed, a sharp decline in the value of $S_t$ may prevent the bank from spending $W_{Min}$, as this might require exceeding daily trading limits or, in an extreme scenario, purchasing all outstanding shares.

\vspace{0.1 cm}
In our simulations, this may result in a few pathological paths where the condition is not satisfied. In practice, many of these cases correspond to Market Disruption Events, which are typically covered by dedicated contractual clauses. The remaining cases, which are rare under realistic market conditions, can be handled through penalty terms in the optimization problem, as detailed below.

\vspace{0.1 cm}
Regarding the hedging portfolio, we follow the same framework described in \cite{Buehler7}. The portfolio is self-financing, and the agent is allowed to trade the firm's shares to mitigate its exposure. In particular, at time $t_n$, the value of the holdings is given by $h_n \cdot S_n$. However, $PnL^{Hedge}_{n}$ must account for all the cash flows required to rebalance the position in $S_n$. At the final settlement $\tau$, it is expressed as:
\begin{equation}
PnL^{Hedge}_{\tau} = \sum_{k = 1}^{n^*} h_k (S_{k+1} - S_k).
\end{equation}

\subsection{Control Problem Optimization}
In this section, we address the control problem optimization framework based on risk measures, described as functions $\rho: \chi \to \mathbb{R}$, where $\chi: \Omega \to \mathbb{R}$ denotes the space of random variables in which $PnL^{ASR}_{\tau}$, $PnL^{Hedge}_{\tau}$ and $PnL_{\tau}$ are defined. Specifically, we rely on the concept of optimized certainty equivalent, or OCE, risk measures \cite{BenTal15, Buehler7}, which allow for efficient computations, as detailed below.

\vspace{0.1 cm}
The optimal policy is obtained as the solution of the following control problem \cite{Buehler7}, which consists of minimizing the chosen risk measure:
\begin{equation}
\pi(PnL_{\tau})  = \underset{\mathcal{P}}\inf \text{ }\rho(PnL_{\tau}) = \underset{\mathcal{P}}\inf \text{ }\rho(PnL^{ASR}_{\tau} + PnL^{Hedge}_{\tau}),
\end{equation}
$
\text{where } \mathcal{P} = \{ n^* \in \mathcal{E}, \text{ } (h_n)_{n \in \{ 1, ..., n^*\}}, \text{ } (b_n)_{n \in \{ 1, ..., n^*\}} \}
.$

\vspace{0.1 cm}
In practice, the objective is to identify the optimal values of $n^*$, $(b_n)_{n \in \{ 1, ..., n^*\}}$, $(h_n)_{n \in \{ 1, ..., n^*\}}$, where $\mathcal{P}$ denotes the set of admissible policies allowed by the contract. Once the program is terminated, the bank must cease both repurchase and hedging activities, which imposes the following conditions:
$$
b_n = 0 \quad \forall  n > n^* \qquad \text{and} \qquad h_n = 0 \quad \forall \  n \geq n^* 
.$$

\vspace{0.1 cm}
Moreover, since it cannot be guaranteed with certainty that $\exists n \in \mathcal{E}: W_n \in [W_{Min}, W_{Max}]$ due to a possible sharp decrease in value of $S_t$, we add a quadratic penalty term $\Psi_{Min}$ to the objective function. The parameter $\beta_{min}$ is a coefficient that aligns the order of magnitude of $\Psi_{Min}$ with that of $\rho(PnL_\tau^{ASR})$, enabling an effective penalization. Once defined $\Psi_{Min} = ((W_{Min}-W_\tau)^+)^2$, we obtain
\begin{equation}
\pi(PnL_\tau) = \underset{\mathcal{P}}\inf \{\rho(PnL_\tau) + \beta_{min} \cdot \Psi_{Min} \}. 
\end{equation}

\vspace{0.1 cm}
We anticipate that in our implementation no penalty is required for exceeding $W_{Max}$. Indeed, in the buyback execution strategies detailed in the Section \ref{sec:buyback}, repurchases automatically cease once $W_{Max}$ is reached.

\vspace{0.1 cm}
Our approach consists in expressing these quantities as parametric functions and subsequently optimizing them. In particular, following \cite{Buehler7}, $h_n = f_{\theta}(\mathcal{S}_n)$, where $f_{\theta}$ is a neural network and $\theta$ denotes its parameters. Analogously, $b_n = f_{\phi}(\mathcal{S}_n)$, where $f_{\phi}$ is another parametric function. The domain of the control problem is then translated from $\mathcal{P}$ to $\mathcal{P'}$:
$$
\mathcal{P'} = \{ n^* \in \mathcal{E}, \text{ } (h_n)_{n \in \{ 1, ..., n^*\}}, \text{ } (b_n)_{n \in \{ 1, ..., n^*\}} : h_n=f_{\theta}(\mathcal{S}_n), \text{ } b_n = f_{\phi}(\mathcal{S}_n) \ \forall n \in \{ 1, ..., n^*\} \}
.$$

\vspace{0.1 cm}
Hence, the vector $\eta = [\theta; \text{ } \phi]$ and $n^*$ fully characterizes the agent’s policy, and the optimization over $\mathcal{P'}$ can therefore be reformulated as an optimization problem over $[n^*, \text{ } \theta, \text{ } \phi]$:
\begin{equation}
\pi(PnL_{\tau})  = \underset{\mathcal{P'}} \inf \text{ }\rho(PnL_{\tau} + \beta_{min} \cdot \Psi_{Min}) = \underset{n^*, \text{ }\theta, \text{ } \phi} \inf \text{ }\rho(PnL_{\tau} + \beta_{min} \cdot \Psi_{Min}). 
\end{equation}

\vspace{0.1 cm}
The procedure described in \cite{Buehler7}, where the authors reformulate the search for the optimal policy $(h_n)_{n \in \{ 1, ..., n^*\}}$ as an optimization over the network parameters $\theta$, is extended here also to the ASR management problem. Optimizing the process $(b_n)_{n \in \{ 1, ..., n^*\}} \in \mathbb{R}^{n^*}$ and $\tau \in \mathcal{E}$ is thus transformed into finding the optimal set of parameters $\phi$ that characterize a parametrized heuristic strategy. For the effectiveness of optimizing heuristic policies to deal with ASR, see \cite{Baldacci10}. In addition, this framework can be applied exclusively to the execution of the buyback, as done in Section \ref{sec:buyback}: the optimization is performed only on $n^*$ and $\phi$, and $PnL_\tau$ is replaced by $PnL_\tau^{ASR}$.

\vspace{0.1 cm}
We now introduce the concept of optimized certainty equivalent by providing its definition \cite{Buehler7}.
\noindent
\begin{definition}[Optimized Certainty Equivalent Risk Measure]\label{def:OCE}
Let $l: \mathbb{R} \to \mathbb{R}$ be a loss function that is continuous, non-decreasing and convex, and let $Z \in \chi$. Then,
\begin{equation}
\rho(Z) = \underset{w \in \mathbb{R}} \inf \{ w + \mathbb{E}[l(-w-Z)] \}
\end{equation}
is an OCE risk measure.
\end{definition}

\vspace{0.1 cm}
By inserting the OCE definition, which satisfies the property of the convex risk measure \cite{Buehler7}, into the control problem formulation, we obtain:
\begin{equation}
\label{eq:OCEcontrol}
\pi(PnL_{\tau}) = \underset{\theta, \text{ } \phi}\inf \ \underset{w \in \mathbb{R}} \inf \{ w + \mathbb{E}[l(-w-PnL_{\tau})] \} = \underset{\theta, \text{ } \phi, \text{ } w \in \mathbb{R}}\inf \{ w + \mathbb{E}[l(-w-PnL_{\tau})] \}
.
\end{equation}

\vspace{0.1 cm}
The value of $w$ can be interpreted as a cash injection added to the asset $Z$. Definition \ref{def:OCE} aims to determine the optimal value of $w$ that minimizes the penalty term $w + \mathbb{E}[l(-w-Z)]$, which accounts for both the cash position $w$ and the expected loss associated with holding $Z + w$. 

\vspace{0.1 cm} 
Thanks to the last representation in equation \ref{eq:OCEcontrol}, the adoption of OCE measures does not require solving an additional minimization problem at each training batch. Instead, it introduces one extra parameter that can be naturally incorporated into the back-propagation process with negligible computational overhead.

\vspace{0.1 cm}
In this work, we adopt the OCE version of the expected shortfall. This measure derives from Definition \ref{def:OCE} by setting
$$
l(x) = \frac{1}{1-\alpha} \cdot max(x, \text{ }0).
$$
where $\alpha \in (0, 1)$ is the confidence level \cite{Buehler7, Foellmer11}. The advantage of this formulation is that, during back-propagation, there is no need to explicitly compute the $\alpha$-quantile of the smallest losses. 

\vspace{0.1 cm}
In order to verify the consistency of our results, we also optimize our policies under the following choice of $l$:
$$
l(x) = x + \frac{1}{2} \cdot \gamma \cdot x^2.
$$
In this case, the resulting risk measure coincides with the mean-variance, where $\gamma$ denotes the risk-aversion coefficient for the variance. Indeed, by substituting this definition of  $l$ into the OCE formula, we obtain:
$$
\rho(Z) = \underset{w \in \mathbb{R}} \inf \{ w + \mathbb{E}[l(- w - Z)] \} = \underset{w \in \mathbb{R}} \inf \{ w + \mathbb{E}[-w - Z + \frac{1}{2} \cdot \gamma \cdot (Z+w)^2]\} =
$$
$$
\underset{w \in \mathbb{R}} \inf \{ \mathbb{E}[- Z +\frac{1}{2} \cdot \gamma \cdot (Z+w)^2]\} = -\mathbb{E}[Z] + \frac{1}{2} \cdot \gamma \cdot \underset{w \in \mathbb{R}} \inf \{ (Z+w)^2]\} = 
$$
$$
= - \mathbb{E}[Z] + \frac{1}{2} \cdot \gamma \cdot \mathrm{Var}(Z)  
.$$

\vspace{0.1 cm}
However, since $l(x)$ is not a decreasing function, $\rho(PnL_\tau)$ does not correspond to an OCE in the formal sense. Specifically, the resulting risk measure does not satisfy the properties of a \textit{convex risk measure}, although the optimization can still be carried out. We note that the mean-variance approach is also employed in \cite{Gueant6}, where it is computed directly from the sample moments of the profit and loss.

\vspace{0.1 cm}
Finally, for numerical stability, the payoff $PnL_{\tau}$ is normalized by $W_{Min}$, which prevents the values from exploding, since the notional amounts involved can be very large.

\subsection{Problem Relaxation}
In practice, in the machine learning framework we develop, the risk measure acts as the loss function and is used to evaluate and update policies. However, to implement neural networks in a unified workflow for the management and hedging of ASR, we require the loss computation to be fully differentiable with respect to the input, namely $\mathcal{S}_t$. This allows us to rely on standard optimization techniques for networks such as gradient descent.

\vspace{0.1 cm}
Once $t_n \in [t_{Min}, \text{ } t_{Max}]$, the agent must decide whether to close or continue the repurchase. This optionality can be represented by a function of the state space $\mathcal{S}_n$ with values in $\{ 0,1 \}$. As shown in \cite{Gueant6}, this feature constitutes the main obstacle to the gradient flow.

\vspace{0.1 cm}
The exercise policy is made differentiable through a stochastic stopping policy $(p_n)_{n \in \mathcal{T}}$, which represents the probability of exercise conditional on $\mathcal{F}_n$. It takes values in $[0, \text{ }1]$ and $p_n = 0 \ \text{ } \forall  n \in \mathcal{T} \setminus \mathcal{E}$.

\vspace{0.1 cm}
The authors link the process $(p_n)_{n \in \mathcal{T}}$ to an effective stopping time $\tau$ by introducing an extended $\sigma\text{-algebra } \mathcal{G} \supseteq \mathcal{F}_{N_{Max}}$ and the probability space $(\Omega, \text{ }\mathbb{P}, \text{ }\mathcal{G})$, which supports a family of $i.i.d$ random variables $(\zeta_n)_{n \in \mathcal{E}}$ uniform in $[0, 1]$ independent of $\mathcal{F}_{N_{Max}}$. 

\vspace{0.1 cm}
The stopping time $\tau = n^* \cdot \delta t$ is defined as $
n^* = \min \{ n \in \mathcal{E} \text{ } | \text{ } \zeta_n \leq p_n \},$ and the exercise decision is given by $\hat{p}_n = \mathbf{1}_{\zeta_n \leq p_n}$, which follows a Bernoulli distribution with parameter $p_n$, conditional on $\mathcal{F}_n$. Hence, the payoff can be expressed as
\begin{equation}
PnL_{\tau} = \sum_{n = N_{Min}}^{N_{Max}} \prod_{k = N_{Min}}^{n - 1}(1-\hat{p}_k) \cdot \hat{p}_n \cdot PnL_n
,
\end{equation}
where $\tau$ depends on the sampling of $(\zeta_n)_{n \in \mathcal{E}}$. At each date $t_n \in [T_{Min}, \text{ } T_{Max}]$ is thus associated an unconditional probability of terminating the program equal to $\tilde{p}_n = \prod_{k = N_{Min}}^{n - 1}(1-p_k) \cdot p_n$, with $\sum_{n = N_{Min}}^{N_{Max}} \tilde{p}_n = 1$.

\vspace{0.1 cm}
Consequently, differentiability is ensured by weighting the loss term $l(-w-PnL_{n})$ and the penalty $\Psi_{Min}$ with their unconditional probability. Therefore, the control problem becomes:
\begin{equation}
\label{eq:controlProblem}
\pi(PnL_{\tau}) = \underset{\theta,\ \phi,\ w}\inf \left\{ 
w + \mathbb{E}\!\left[
\sum_{n=N_{\min}}^{N_{\max}}
\tilde{p}_n \cdot
\left( l(-w-PnL_n) + \beta_{min} \cdot ((W_{\min}-W_n)^+)^2 \right)
\right]
\right\}.
\end{equation}

\subsection{ASR Pricing}
In conclusion, we address the pricing of ASR contracts. In a complete market, the risk-neutral valuation formula determines the contract price $\Pi^{ASR}(t_0)$, which represents the value of the replicating portfolio at time $t_0$. However, this portfolio cannot always be replicated in practice, since it is defined in terms of Greeks that, as discussed in the Introduction, suffer from several limitations and may be infeasible to implement due to the constraints imposed by the term sheet or market regulations. Consequently, the theoretical value $\Pi^{ASR}(t_0)$ may lose its practical relevance.

\vspace{0.1 cm}
To clarify this point, we focus on the discount $\delta$, which is the standard reference in the industry. In the competitive ASR market, banks secure contracts by offering a greater discount than their competitors; as a result, practitioners commonly refer to this discount as the “price” of the buyback program. In practice, the better the performance in the execution, and therefore the higher the value generated by the ASR, the larger the discount that can be offered. Typically, offering a higher discount is made possible either by enriching the term sheet with additional features, such as a \textit{greenshoe} on the notional amount, or by modifying the payoff structure itself.

\vspace{0.1 cm}
If replication through Greeks were possible, the bank would offer the so-called \textit{fair discount} $\delta^{fair}$, obtained by requiring the contract to be fair. In our settings with interest rates set to zero, the fair discount in the risk-neutral valuation framework corresponds to the value of $\delta$ that, substituted into $PnL^{ASR}_{\tau}$, satisfies $\mathbb{E}^{\mathbb{Q}} [PnL^{ASR}_{\tau}] = 0$
where $\mathbb{Q}$ is a risk-neutral measure. It follows:
\begin{equation}
\delta^{fair} = 1 - \frac{\mathbb{E}^{\mathbb{Q}} [W_{n^*}] }{\mathbb{E}^{\mathbb{Q}} [A_{n^*} \cdot Q_{n^*}] }.
\end{equation}

\vspace{0.1 cm}
The definition of $\delta^{fair}$ and the ones that follow hold also in the relaxed settings, where $PnL_{\tau}, \text{ or } PnL^{ASR}_{\tau}$ in Section \ref{sec:buyback}, are distributed over the exercise time window:
\begin{equation}
PnL = \sum_{n = N_{Min}}^{N_{Max}} \prod_{k = N_{Min}}^{n - 1}(1-p_k) \cdot p_n \cdot PnL_n.
\end{equation}

\vspace{0.1 cm}
However, since it is not possible to completely eliminate risk, the bank cannot offer the fair discount to its client. An alternative approach is provided by the concept of indifference pricing. Due to the cash-invariance property of convex risk measures, we have:
$$
\rho(PnL_{\tau} + \rho(PnL_{\tau})) = \rho(PnL_{\tau}) - \rho(PnL_{\tau}) = 0.
$$

\vspace{0.1 cm}
In particular, $\rho(PnL_{\tau})$ can be interpreted as the indifference price, that is, the amount of cash required to make the position acceptable according to the risk measure $\rho$ \cite{Buehler7}. Since it is typically negative for ASR contracts, as shown in the following Sections, the bank recognizes an intrinsic value in these products. However, rather than transferring $|\rho(PnL_{\tau})|$ directly to its client, the bank returns this value in the form of a discount. Consequently, we introduce the concept of the \textit{indifference discount}, defined as the value $\delta^{ind}$ such that $\rho(PnL_{\tau}) = 0$.

\vspace{0.1 cm}
The quantities $\delta^{fair}$ and $\delta^{ind}$ provide valuable insights on the performance of our implementation. Specifically, $\delta^{fair}$ denotes the discount that assumes perfect hedging, while $\delta^{ind}$ represents the discount that can be offered given the actual hedging capabilities. It is worth noting that the definition of $\delta^{fair}$ involves $PnL^{ASR}_{\tau}$: once the ASR management strategy is fixed, it measures the discount enabled by that policy. Conversely, $\delta^{ind}$ is obtained by considering the overall profit and loss, $PnL_{\tau} = PnL_{\tau}^{ASR} + PnL_{\tau}^{Hedge}$, and therefore also reflects the hedging performance. In our numerical analysis, both quantities are estimated on an independent test set. For the fair discount, we approximate the expected values under $\mathbb{Q}$ by their sample averages, as in our settings $\mathbb{P} = \mathbb{Q}$.

\section{ ASR Management}
\label{sec:buyback}

\subsection{Smooth Bang-Bang Strategy}
We start our analysis by addressing the execution of the buyback. In this section, we describe the \textit{smooth bang-bang} policy, which plays the role of the benchmark implementation to be compared to neural network models.

\vspace{0.1 cm}
It is based on two almost linear repurchase regimes: the first targeting $W_{Min}$ at the last termination date $N_{Max}$, while the second $W_{Max}$ at the minimum termination date $N_{Min}$.
\[
\left\{
\begin{aligned}
  &q_n^{min} = \max \left(\min\left(\underline{\nu}_n, \text{ } \left(W_{Min} - W_{n-1}\right)^+\right), \text{ } \frac{W_{Min} - W_{n-1}}{S_n \cdot (N_{Max} - n + 1)}\right)\\[0.5em]
  &q_n^{max} = \min\left(\overline{\nu}_n, \text{ } \frac{W_{Max} - W_{n-1}}{S_n \cdot \max(1, N_{Min} - n + 1)}\right)
\end{aligned}
\right.
\]
The repurchase process $(b_n)_{n \in \mathcal{T}}$ is defined as:
\begin{equation}
b_n(S_n, \text{ } A_n) = \min\left(q_n^{max}, \text{ } \max\left(q_n^{min}, \text{ } b_n^*(S_n, \text{ } A_n)\right)\right) \quad \forall  n \in \mathcal{T},
\end{equation}
$\text{where} \quad b_n^*(S_n, \text{ } A_n) = q_n^{max} + \frac{q_n^{min} - q_n^{max}}{\delta_{r}} \cdot \left(\frac{S_n}{A_n} - (1 + \epsilon_{r}) + \frac{\delta_{r}}{2}\right).$

\begin{figure}[H] 
    \centering
    \includegraphics[width=0.9\textwidth]{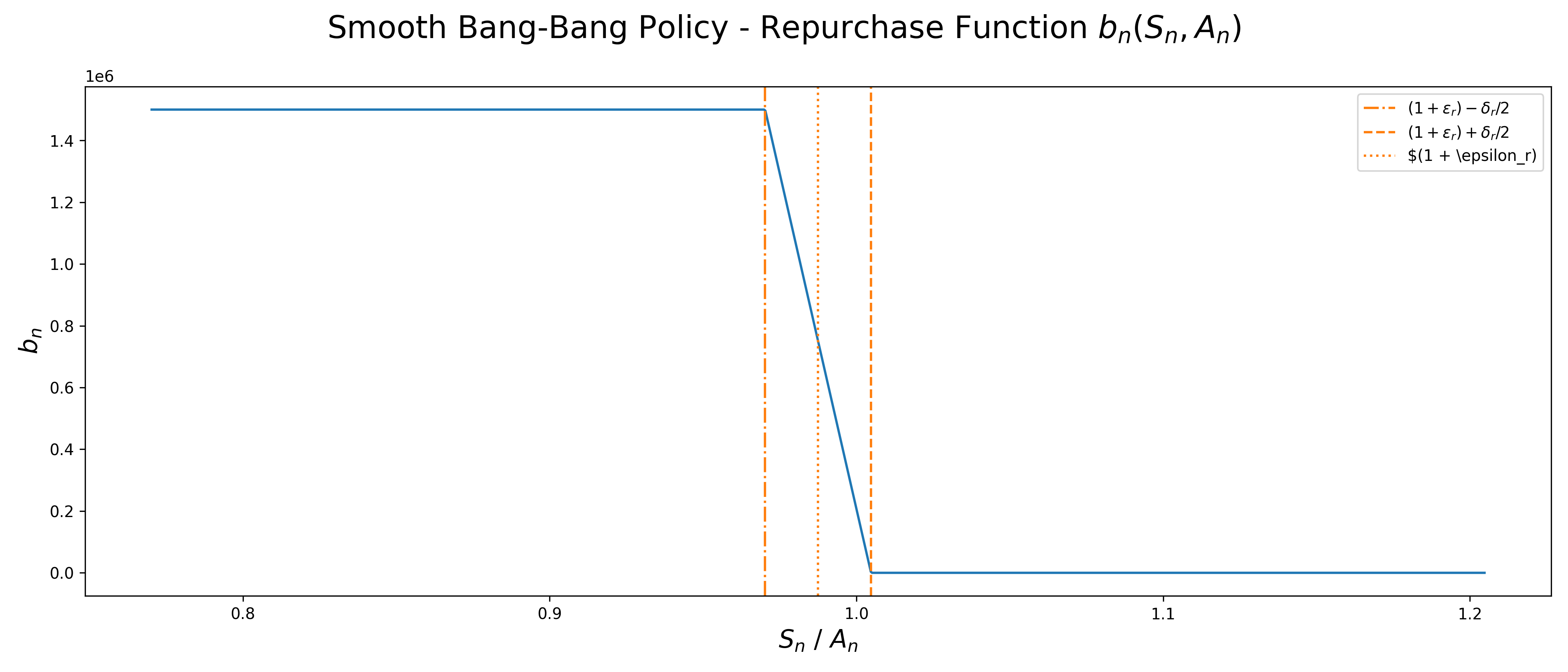} 
    \caption{$b_n(S_n, \text{ } A_n)$ function: when the ratio between $S_n \text{ and } A_n$ falls in the interval $[1 + \epsilon_r - 0.5 \cdot \delta_r, \text{ } 1 + \epsilon_r + 0.5 \cdot \delta_r]$, the repurchase $b_n$ is determined by a linear interpolation over $[q_n^{min}, \text{ } q_n^{max}]$.} 
    \label{fig:smoothedStep} 
\end{figure}

\vspace{0.1 cm}
The number of shares repurchased at $t_n$ can take any value value between the two regimes and depends on $S_n$ and $A_n$ solely through their ratio, centered at $1 + \epsilon_r$, while the parameter $\delta_r$ controls the slope, as shown in figure \ref{fig:smoothedStep}.

The intuition behind this approach is straightforward: when $S_n$ falls below $A_n$, it becomes advantageous to accelerate the repurchase, as the underlying asset is dragging down the benchmark value. Moreover, the bank is repurchasing shares at a price lower than the potential compensation, thereby increasing its profit. 

\vspace{0.1 cm}
As described in Section \ref{sec:model}, the hedging framework detailed in \cite{Buehler7} requires the flow of gradient. We thus introduce probability process $(p_n)_{n \in \mathcal{E}}$, shifting the problem to the relaxed settings. Furthermore, the definition of $p_n$ must ensure that $W_{n^*} \in [W_{Min}, W_{Max}]$ when the exercise time $\tau$ is sampled using the variables $(\zeta_n)_{n \in \mathcal{E}}$. In the relaxed setting, this requirement becomes $W_n \in [W_{Min}, \text{ } W_{Max}] \quad \forall n \in \mathcal{E}: \prod_{k = N_{Min}}^{n - 1}(1-p_k) \cdot p_n > 0.$
For this reason, the conditional probability is expressed as a function of $W_n$:
\begin{equation}
p_n(W_n) = \max\left(0, \text{ } \min\left(1, \text{ } \frac{1}{\delta_p} \cdot \left(W_n^* - \epsilon_p + \frac{\delta_{p}}{2}\right)\right)\right) \quad \forall n \in \mathcal{E},
\end{equation}
where $W_n^* = \left(W_n - \frac{1}{2} \cdot (W_{Min} + W_{Max})\right) / \left(\frac{1}{2} \cdot (W_{Max} - W_{Min})\right)$, mapping $[W_{Min}, W_{Max}] \text{ into } [-1,1].$

\vspace{0.1 cm}
The value of $p_n$ is centered at $\epsilon_p$ and linearly interpolated over the interval $I_{\mathcal{E}} = [\epsilon_p - \frac{1}{2} \delta_p, \text{ } \epsilon_p + \frac{1}{2} \delta_p]$, with the result capped within $[0,1]$. The slope is determined by the parameter $\delta_p$, and at $T_{Max}$ the agent is forced to close the program. If $I_{\mathcal{E}} \subseteq [-1,1]$, the formulation of $p_n$ guarantees a non-zero probability of exercise only when $W_{n} \in [W_{Min}, W_{Max}]$. For this strategy, the set of trainable parameter is $\phi = [\epsilon_r, \text{ } \delta_r,\text{ } \epsilon_p, \text{ } \delta_p]$.

\subsection{The network strategy}
In this section, we address what we refer to as the \textit{network} strategy. The relationship between the processes $(b_n)_{n \in \mathcal{T}}$, $(p_n)_{n \in \mathcal{E}}$ and the full state space $(\mathcal{S}_n)_{n \in \mathcal{T}}$ is delegated to a neural network, represented by a function $f_{\phi}: \mathbb{R}^3 \to [0,1]^2$, with a sigmoid activation function in the output layer. At each time step $t_n$, the network computes
\begin{equation}
\label{eq:networkPolicy}
[b_n^*, \text{ } p_n] = f_{\phi}\left(\frac{n-N_{Min}}{N_{Min}} \text{ }\frac{S_n}{A_n}, \text{ } \frac{W_{n-1}}{W_{Min}}\right),
\end{equation}
with $b_n = v_n^{max} + \left(v_n^{min} - v_n^{max}\right) \cdot b_n^* \quad \forall n \in \mathcal{T}$. The quantities $v_n^{min}$ and $v_n^{max}$ are defined in the following.

\vspace{0.1 cm}
The inputs of the network are rescaled to have comparable magnitudes, a step that proves crucial for successful training. Furthermore, they are dimensionless, as in \cite{Gueant6}.

\vspace{0.1 cm}
Our experiments suggest that the total number of shares already repurchased $Q_{n-1}$ does not contribute significantly once the other state variables in $\mathcal{S}_n$ are taken into account. Similarly, it is sufficient to represent the underlying value $S_n$ and its running average $A_n$ through their ratio, as including them separately does not offer additional benefit. The number of inputs is thus reduced from five to three, decreasing the model complexity. 

\vspace{0.1 cm}
Before introducing the expressions for $v_n^{min}$ and $v_n^{max}$, which provide greater flexibility in order to exploit the information contained in the state space, we define the quantity $(d_n)_{n \in \mathcal{T} \setminus { N_{Max}}}$. At time $t_n$, $d_n$ denotes the number of days required to reach $W_{Min}$ when purchasing the maximum number of shares $\overline{\nu}_k$ for all $k > n$ up to maturity:
\begin{equation}
d_n = \lceil d_n^* \rceil, \quad \text{ with } \quad d_n^* = \frac{W_{Min} - W_{n-1}}{\overline{\nu}_n \cdot S_n^*}.
\end{equation}
The term $S_n^*$ acts as a buffer against a sharp decline in the spot price. Within the Black--Scholes framework employed in our analysis, a possible choice for $S_n^*$ is
\begin{equation}
S_n^* = S_n \cdot e^{- \frac{1}{2} \sigma^2  dt \ +\  \Phi^{-1}(\gamma^*) \cdot \sigma \sqrt{dt}},
\end{equation}
where $\Phi(x)$ denotes the cumulative distribution function of a standard normal random variable. In practice, $S_n^*$ represents the worst-case projection of the asset value $S_{n+1}$, which we compute at a confidence level of $\gamma^* = 0.05$.

\vspace{0.1 cm}
Finally, we detail the definitions of $v_n^{min} \text{ and } v_n^{max}$. In the next section, we show their contribution to the increase of performance and how they would impact the \textit{smooth bang-bang}. For $n \in [1, \text{ } N_{Max} - 1]$:
\begin{equation}
v_n^{min} = \quad
\left\{
\begin{aligned}
  & \underline{\nu}_n,  &&\text{if } \quad d_n < N_{Max} - n\\[0.5em]
  & \min(\overline{\nu}_n, \text{ } \frac{W_{Min} - W_{n-1} - \overline{\nu}_n \cdot (N_{Max} - n) \cdot S^*_n}{S_n}),  &&\text{if } \quad d_n \geq N_{Max} - n
\end{aligned}
\right.
.
\end{equation}
At maturity $v_{N_{Max}}^{min} = \max\left(\underline{\nu}_n, \text{ } \frac{W_{Min} - W_{n-1}}{S_n}\right).$

\vspace{0.1 cm}
Unlike the previous policy, the lower bound for daily repurchases is not defined as an even allocation aimed at reaching $W_{Min}$ by $T_{Max}$, since such a rule would enforce purchases even when they are not advantageous. Instead, the policy allows the bank to make the minimum admissible purchase $\underline{\nu}_n$ on most days. At the same time, it continuously monitors the minimum number of days required to reach $W_{Min}$. If the remaining time is insufficient to reach the minimum notional amount at the current repurchase rate, the policy increases daily purchases accordingly to ensure that the constraint is met by maturity.

\vspace{0.1 cm}
The definition of $v_n^{max}$ closely follows that of the \textit{bang-bang}. For $n \in [1, \text{ } N_{Min} - 1]$:
\begin{equation}
v_n^{max} = \quad
\left\{
\begin{aligned}
  & \min(\overline{\nu}_n, \text{ } \frac{W_{Max}-W_{n-1}}{S_n \cdot (N_{Min}-n+1)}),  &&\text{if } \quad d_n < N_{Max}- n\\[0.5em]
  & \max(v_n^{min}, \text{ } \frac{W_{Max}-W_{n-1}}{S_n \cdot (N_{Min}-n+1)}),  &&\text{if } \quad d_n \geq N_{Max}- n
\end{aligned}
\right.
.
\end{equation}
For $n \in [T_{Min}, \text{ } T_{Max}]$, the bank may reach $W_{Max}$ in a single day:
$$
v_n^{max} = \min\left(\overline{\nu}_n, \text{ } \frac{W_{Max}-W_{n-1}}{S_n}\right).
$$

\vspace{0.1 cm}
We now turn to explain how the constraints on the notional are met. We note that, by construction, the \textit{network} strategy ensures that the notional value $W_t$ reaches $W_{Min}$ before maturity, except the cases where the underlying asset $S_t$ experiences a sharp decline, while also preventing $W_t$ from exceeding $W_{Max}$. Nevertheless, the implementation may still assign a non-zero probability of termination when $W_n \notin [W_{Min}, W_{Max}]$, thereby violating the contract terms. This is avoided by checking, for each $n \in \mathcal{E}$, whether $W_n \in [W_{Min}, W_{Max}]$. If the notional constraints condition is not met, $p_n$ is retrospectively set to 0. 

\vspace{0.1 cm}
However, the bank must terminate the repurchase. Therefore, we force termination at maturity by setting $p_{N_{Max}} = 1$. Consequently, the term $\Psi_{Min}$ can penalize the cases where $W_{N_{Max}} < W_{Min}$, reducing the scenarios in which $W_n \notin [W_{Min}, W_{Max}] \quad \forall  n \in \mathcal{E}.$

\vspace{0.1 cm}
Although retrospectively imposing $p_n$ to 0 is non-differentiable, thereby excluding certain paths from the backpropagation step, its impact on the overall optimization is limited. Indeed, the interaction between the policy design, for which the notional usually falls within the target interval, the penalty term $\Psi_{Min}$ for extreme scenarios and the fulfillment of the requirement $\sum_{n = N_{Min}}^{N_{Max}} \tilde{p}_n = 1$ allows the network to quickly learn the optimal timing for contract closure.

\vspace{0.1 cm}
In our implementation, the nearly enforcement of the notional constraints reduces the impact of the penalty term $\Psi_{Min}$ and simplifies the calibration of the parameter $\beta_{min}$, whose optimal value depends on the specific characteristics of the contract under analysis and could therefore limit the generalizability of the approach. As shown in the next section, our implementation strikes a balance between the freedom granted to the agent and the guidance of its actions, contributing to a faster and more stable training process.

\subsection{Smooth Bang-Bang vs Network}
We begin by introducing the model and the contract used throughout the work. As discussed above, the price of the underlying share is assumed to follow the Black--Scholes dynamics. Nevertheless, our framework can be applied under any specification adopted for the underlying asset. As discussed in Section \ref{sec:model}, interest rates are neglected by setting $r = 0$.
\begin{table}[H]
\centering
\begin{tabular}{|c|c|c|c|}
\hline
$S_0$ [\euro] & $\sigma_{year}$ & r \\
\hline
$45$ & $0.21$ & $0$ \\
\hline
\end{tabular}
\caption{Black--Scholes parameters.}
\label{tab:blackParameters}
\end{table}

The term sheet of the ASR contract is presented in table \ref{tab:contract}:
\begin{table}[H]
\centering
\begin{tabular}{|c|c|c|c|c|c|c|}
\hline
$T_{Min}$ [day]& $T_{Max}$ [day]& $W_{Min}$ [\euro]& $W_{Max}$ [\euro]& $\delta$ & $\underline{\nu}_n$ & $\overline{\nu}_n$  \\
\hline
$22$ & $63$ & $810 \cdot 10^6$ & $990 \cdot 10^6$ & $0$ & $0$ & $1.5 \cdot 10^6$ \\
\hline
\end{tabular}
\caption{ASR term sheet. The values of $\underline{\nu}_n \text{ and } \overline{\nu}_n$ are constant $\forall n \in \mathcal{T}$.}
\label{tab:contract}
\end{table}

As mentioned in Section \ref{sec:model}, we rely on two risk measures: the expected shortfall, which focuses on the left tail of the $PnL$ distribution, and the mean-variance loss, which provides a symmetric assessment of risk. To keep the presentation simple and clear, for $\delta^{fair}$ and $\delta^{ind}$, we report only the values obtained under the expected shortfall optimization. The penalty coefficient of $\Psi_{Min}$ is set to $\beta_{min} = 500$.
\begin{table}[H]
\centering
\begin{tabular}{|c|c|c|c|}
\hline
Risk Measure & Symbol & Parameter \\
\hline
Expected Shortfall     & $ES_{0.75}$ & $\alpha = 0.75$ \\
\hline
Mean-Variance  & $MV_{2.5 \cdot 10^2}$ & $\gamma = 2.5 \cdot 10^2$ \\
\hline
\end{tabular}
\caption{Specification of the risk measures. Both risk measures are computed using the $PnL$ normalized by $W_{\min}$.}
\label{tab:riskMeasureTab}
\end{table}

\vspace{0.1 cm}
For all the numerical results, we employ the standard machine learning data split into training, validation, and test sets. Specifically, the payoff distributions and evaluation metrics, namely the risk measure and the discounts $\delta^{fair}$ and $\delta^{ind}$, are computed on the test set. 

\vspace{0.1 cm}
Regarding the \textit{network} strategy, $f_{\phi}$ consists of four layers: the first three contain 128 neurons each, followed by ReLU activation functions, while the final layer has two neurons with a sigmoid activation. Normalization layers are not used because the key factor for successful training appears to be the scaling of the inputs. 

\vspace{0.1 cm}
We first qualitatively assess the performance of the \textit{smooth bang-bang} and \textit{network} strategy with the payoff distributions. All the distributions of Section \ref{sec:buyback} and Section \ref{sec:hedging} are obtained by optimizing the expected shortfall, and the mean-variance risk measure yields  similar outcomes. The figures are obtained on a test set of 3000 cases. Moreover, also the discounts $\delta^{fair} \text{ and } \delta^{ind}$ come from the expected shortfall optimization.
\begin{figure}[H] 
    \centering
    \includegraphics[width=0.9\textwidth]{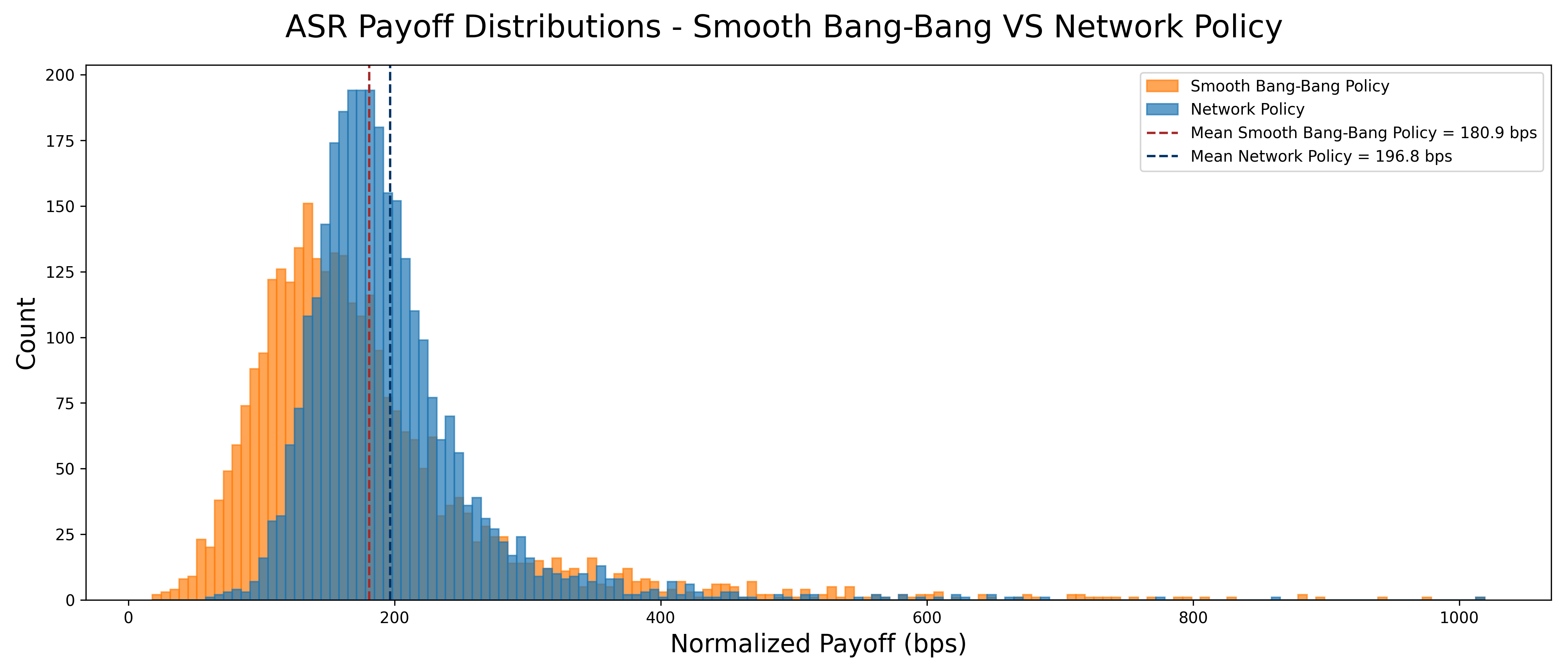} 
    \caption{Payoff distribution for the \textit{smooth bang-bang} and \textit{network} strategy. The payoff is normalized by $W_{Min}$ and expressed in basis points. } 
    \label{fig:buybackSmoothVSNetwork} 
\end{figure}

\vspace{0.1 cm}
The \textit{network} model exhibits a more concentrated payoff distribution with a thinner left tail, stabilizing performance, as adverse scenarios are less frequent compared to the \textit{smooth bang-bang}. Moreover, the higher expected values observed under the \textit{network} policy, also consistent with the mean-variance optimization, suggest an enhanced ability to manage the repurchase process. These findings are further supported by the numerical results in Table \ref{tab:tabSection3}: 
\begin{table}[H]
\centering
\begin{tabular}{|c|c|c|c|c|c|}
\hline
Strategy & $ES_{0.75}$ & $MV_{2.5 \cdot 10^2}$ & $\delta^{fair}$ & $\delta^{ind}$ \\
\hline
\textit{Smooth Bang-Bang} & $-94.0$  & $-62.8$ & $156.5$ & $82.1$ \\
\hline
\textit{Network}          & $-136.8$ & $-161.1$ & $167.4$ & $114.6$ \\
\hline
\end{tabular}
\caption{Risk measures and discount for \textit{smooth bang-bang} and \textit{network} policies, expressed in basis points.}
\label{tab:tabSection3}
\end{table}

\vspace{0.1 cm}
We observe that the expected shortfall takes negative values, which is consistent with the strictly positive payoff distributions. By definition, the expected shortfall represents the average loss in the $1-\alpha$ worst-case scenarios. Therefore, a positive value indicates an actual loss, which never occurs. Consequently, when the metric takes negative values, a larger absolute value corresponds to a larger profit rather than a larger loss. This same holds for the mean-variance criterion: considering negative outcomes, a larger value in absolute terms still indicates a more favorable outcome. Indeed, the optimization is formulated as a minimization problem. Hence the reported values correspond to
$$
- \mathbb{E}[PnL] + \frac{1}{2} \cdot \gamma \cdot \mathrm{Var}(PnL) = - (\mathbb{E}[PnL] - \frac{1}{2} \cdot \gamma \cdot \mathrm{Var}(PnL))
.$$

\vspace{0.1 cm} 
The \textit{network} policy exhibits a larger absolute value for both metrics. In other words, both evaluation criteria consistently indicate that the \textit{network} implementation outperforms the \textit{smooth bang-bang} policy. Moreover, the higher fair discount, reflecting the greater expected value of the payoff distribution, implies that the former policy generates more value while managing the ASR.

\vspace{0.1 cm}
Furthermore, the network policy also exhibits a higher indifference discount $\delta^{ind}$, which incorporates exposure assessment through the measure $\rho$. This implies that, even accounting for risk, the bank can still offer a larger discount, providing an alternative to standard pricing that better reflects operational capabilities. 

\vspace{0.1 cm}
We now turn to investigating the factors driving this performance improvement. In a manner similar to \cite{Gueant6}, where networks are used to introduce a modification from a simpler policy, the \textit{network} strategy is obtained by improving the \textit{smooth bang-bang}. The main differences between the two policies are three: 
\begin{enumerate}[noitemsep, topsep=0pt]
    \item the use of a neural network to compute $b_n$
    \item the use of a neural network to compute $p_n$
    \item the definition of the interval $[v_n^{min}, v_n^{max}]$, which allows for greater flexibility in market repurchases compared to the bounds $b_n \in [q_n^{min}, q_n^{max}]$ used in the \textit{smooth bang-bang} policy.
\end{enumerate}

\vspace{0.1 cm}
\noindent
We address this by designing the following configurations, each derived from the smooth bang-bang policy by modifying one of the aspects listed above. Configuration C, however, involves changes in two aspects.
\begin{itemize}[noitemsep, topsep=0pt]
\item Configuration A: the neural network is used to compute $b_n$ only
\item Configuration B: the neural network is used to compute $p_n$ only
\item Configuration C: $b_n$ remains in $[q_n^{min}, q_n^{max}]$ as in the \textit{smooth bang-bang}, while both $b_n$ and $p_n$ are computed according to the implementation \ref{eq:networkPolicy} of the \textit{network} policy.
\item Configuration D: $b_n$ and $p_n$ are computed using the piecewise linear function as in the \textit{smooth bang-bang}, but $b_n \in [v_n^{min}, v_n^{max}]$
\end{itemize}

\vspace{0.1 cm}
Table \ref{tab:networkConfigurations} summarizes the characteristics of the configurations and presents the results they achieve. For clarity, a dash (-) indicates that the feature in the column is implemented as in the \textit{smooth bang-bang} policy, while $f_{\phi}$ denotes the use of a neural network.
\begin{table}[H]
\centering
\begin{tabular}{|c|c|c|c|c|c|c|c|}
\hline
Strategy & $b_n$ & $p_n$ & $b_n \in$ & $ES_{0.75}$ & $\delta^{fair}$ & $\delta^{ind}$\\
\hline
\textit{Smooth Bang-Bang} & -          & -          & $[b_n^{min}, \text{ } b_n^{max}]$ & $-94.0$ & $156.5$ & $82.1$ \\
\hline
\textit{Network}      & $f_{\phi}$ & $f_{\phi}$ & $[v_n^{min}, \text{ } v_n^{max}]$ & $-136.8$ & $167.4$ & $114.6$ \\
\hline
Config. A             & $f_{\phi}$ & -          & $[b_n^{min}, \text{ } b_n^{max}]$ & $-103.3$ & $156.3$ & $87.7$ \\
\hline
Config. B             & -          & $f_{\phi}$ & $[b_n^{min}, \text{ } b_n^{max}]$ & $-97.8$ & $161.4$ & $83.7$ \\
\hline
Config. C             & $f_{\phi}$ & $f_{\phi}$ &  $[b_n^{min}, \text{ } b_n^{max}]$& $-105.5$ & $156.7$ & $89.3$ \\
\hline
Config. D             & -          & -          & $[v_n^{min}, \text{ } v_n^{max}]$ & $-47.3$ & $133.0$ & $40.2$ \\
\hline
\end{tabular}
\caption{Expected shortfall and benchmark discounts for configurations A–D, expressed in basis points. A dash (-) indicates that the feature in the column is implemented as in the \textit{smooth bang-bang}, using a piecewise linear function. Neural networks are indicated by $f_{\phi}$, whose architecture is the one adopted for the \textit{network} policy, except for the last layer for configurations A and B, which has only one neuron.}
\label{tab:networkConfigurations}
\end{table}

\vspace{0.1 cm}
Table \ref{tab:networkConfigurations} suggests that the performance improvement arises from the interaction of multiple components rather than from any single modification. Among the configurations, using a neural network to compute $b_n$, configuration A, produces the largest individual effect. Nevertheless, this enhancement alone is insufficient to match the performance of the \textit{network} policy. When the network is also employed to estimate $p_n$, configuration C, the model shows further improvement, whereas configuration B performs comparably to the \textit{smooth bang-bang} baseline. These results indicate a clear synergistic effect: joint learning of both $b_n$ and $p_n$ yields better outcomes than optimizing either component separately, and the gains of the \textit{network} policy cannot be attributed to a single factor.

\vspace{0.1 cm}
Furthermore, although configuration D performs noticeably worse, redefining $[q_n^{\min}, \text{ } q_n^{\max}]$ into $[v_n^{\min}, \text{ } v_n^{\max}]$ contributes to performance enhancement when the repurchases and exercise probabilities are parametrized by networks, as configuration C does not equal the \textit{network} strategy. This may be explained by the limited information contained in the ratio $\frac{S_n}{A_n}$, which appears to be insufficient to exploit the broader range allowed for $b_n$. The inclusion of additional state variables via neural networks enables the model to fully leverage the expanded flexibility.

\section{ASR Hedging}
\label{sec:hedging}

\subsection{Sequential and Joint Hedging}
We now turn to mitigate the exposure on the underlying by introducing the \textit{sequential} hedging model, whose purpose is to apply the framework proposed in \cite{Buehler7} and to assess whether it remains effective when used for ASR contracts. 

\vspace{0.1 cm}
For this reason, a first simplified approach is fixing the repurchase behavior. The financial institution first optimizes the execution of the repurchase as done in Section \ref{sec:buyback}. Once fixed the optimal set of parameters $\phi$, the second steps consists on training the set of weights $\theta$. 

\vspace{0.1 cm}
However, this approach consists of two optimization steps that are handled independently, leading to potential suboptimal outcomes. In the first stage, the executor of the program determines the optimal repurchase policy by minimizing the risk measure $\rho$, thereby selecting a specific trade-off between profit and risk. In the second stage, however, part of the residual risk is hedged away. As a consequence, the policy chosen in the first step may be overly conservative: the agent might prefer a more aggressive repurchase strategy if it could anticipate that some of the risk would later be mitigated.

\vspace{0.1 cm}
This limitation becomes evident from Figure \ref{fig:buybackSmoothVSNetwork}. The payoff distributions are strictly positive, reflecting the agent’s attempt to minimize losses during the first optimization stage. However, a more permissive policy, which tolerates a heavier left tail, might be optimal once the subsequent hedging step is taken into account, since part of the downside risk would eventually be neutralized.

\vspace{0.1 cm}
These considerations motivate the introduction of the \textit{joint} model, where the execution of the ASR and the hedging strategy are optimized simultaneously, which may lead to a more efficient policy. The bank thus faces the full control problem \eqref{eq:controlProblem}, described in detail in Section \ref{sec:model}.

\vspace{0.1 cm}
We now detail the implementation of the architecture for hedging. The value of $h_n$ is computed by the neural network $f_{\theta}$, which has a linear activation on the last layer, allowing $h_n \in \mathbb{R}$ since no constraints are imposed on the hedging portfolio:
\begin{equation}
h_n = \lambda \cdot f_{\theta}\left(\frac{n - N_{Min}}{N_{Min}}, \text{ } \frac{S_n}{A_n}, \text{ } \frac{W_{n-1}}{W_{Min}}\right) \quad \forall n \in \mathcal{T} \setminus \{ N_{Max}\},
\end{equation}
with $\lambda = (W_{Min} + W_{Max}) / (S_0 \cdot (N_{Min} + N_{Max}))$.

\vspace{0.1 cm}
The non-trainable multiplier $\lambda$ plays a crucial role from a numerical perspective, as it ensures that all quantities, and, more importantly, their gradients, enter the model at a comparable order of magnitude. The volumes involved in ASR transactions are typically very large, which can overshadow variables operating on a smaller scale. In practice, we observed that, without the multiplier $\lambda$, $PnL^{Hedge}$ is initially negligible compared to $PnL^{ASR}$, resulting in negligible gradients. Consequently, during training, the model learns to manage the ASR by optimizing the parameters $\phi$, while the weights $\theta$ remain effectively untrained. This causes the training process to converge to a local minimum equivalent to optimizing only the ASR, as in the previous Section.

\vspace{0.1 cm}
We turn to present the performance of our implementation on the contract described in Section \ref{sec:buyback}. The function $f_{\theta}$ consists of six layers. The first five layers each contain $150$ neurons and are followed by a ReLU activation function, while the final layer has a single neuron.

\vspace{0.1 cm}
In particular, we first analyze the \textit{sequential} model, which proves the effectiveness of the hedging framework for ASR. This is confirmed by the results shown table \ref{tab:sequentialRiskMeasuresTab}: when $PnL^{Hedge}_n$ is included, both the expected shortfall and the mean-variance decrease, demonstrating the ability of the hedging portfolio to mitigate risk.
\begin{table}[H]
\centering
\begin{tabular}{|c|c|c|}
\hline
Model & $ES_{0.75}$ & $MV_{2.5 \cdot 10^2}$ \\
\hline
\textit{Smooth Bang-Bang}            & $-94.0$  & $-62.8$ \\
\hline
\textit{Sequential Smooth Bang-Bang} & $-147.4$ & $-162.2$ \\
\hline
\textit{Network}             & $-136.8$ & $-161.1$ \\
\hline
\textit{Sequential Network}  & $-158.1$ & $-174.5$ \\
\hline
\end{tabular}
\caption{Risk measures for the \textit{sequential} models hedging the \textit{smooth bang-bang} and the \textit{network} policies, expressed in basis points.}
\label{tab:sequentialRiskMeasuresTab}
\end{table}

\vspace{0.1 cm}
The impact of the \textit{sequential} hedging model is more pronounced on the benchmark \textit{smooth bang-bang} policy, whose risk measures show a larger reduction compared to those obtained with the \textit{network} policy. This can be attributed to the fact that, as illustrated in Section \ref{sec:buyback}, the \textit{network} strategies not only increase the contract’s value but also produce a more peaked distribution, thereby already embedding some degree of protection.

\vspace{0.1 cm}
The positive impact of the hedging portfolio is further confirmed by the values of the offered discounts on the benchmark price, which align with the indications provided by the risk measures.
\begin{table}[H]
\centering
\begin{tabular}{|c|c|c|c|}
\hline
Model & $\delta^{fair}$ & $\delta^{ind}$ & $\delta^{fair} - \delta^{ind}$ \\
\hline
\textit{Smooth Bang-Bang}            & $156.5$ & $82.1$ & $74.4$\\
\hline
\textit{Sequential Smooth Bang-Bang} & $156.5$ & $126.5$ & $30.0$ \\
\hline
\textit{Network}                     & $167.4$ & $114.6$ & $52.8$ \\
\hline
\textit{Sequential Network}          & $167.4$ & $132.4$ & $35.0$ \\
\hline
\end{tabular}
\caption{Fair and indifference discounts for the \textit{sequential} models hedging the \textit{smooth bang-bang} and the \textit{network} policies, expressed in basis points.}
\label{tab:sequentialDiscountTab}
\end{table}

The increase in the indifference discount $\delta^{ind}$, for both strategies, confirms that the hedging framework detailed in \cite{Buehler7} can be successfully applied to ASR, producing tangible performance improvements. Although the \textit{network} model still outperforms the \textit{baseline}, the addition of the hedging portfolio narrows the gap $\delta^{fair} - \delta^{ind}$, making it comparable to, and, in fact, slightly better than, that of the \textit{network} policy.

\vspace{0.1 cm}
It is worth noting that the fair discount $\delta^{fair}$ remains unchanged when considering the \textit{sequential} models, as it depends solely on $PnL^{ASR}_\tau$, which is unaffected once the repurchase strategy has been fixed. Consequently, the distributions of $PnL^{ASR}_\tau$ for any given policy and its associated sequential hedging are identical and remain the same as those presented in Section \ref{sec:buyback}. We show in Figure \ref{fig:sequentialFig} the distribution of $PnL_\tau = PnL^{ASR}_\tau + PnL^{Hedge}_\tau$:
\begin{figure}[H] 
    \centering
    \includegraphics[width=0.9\textwidth]{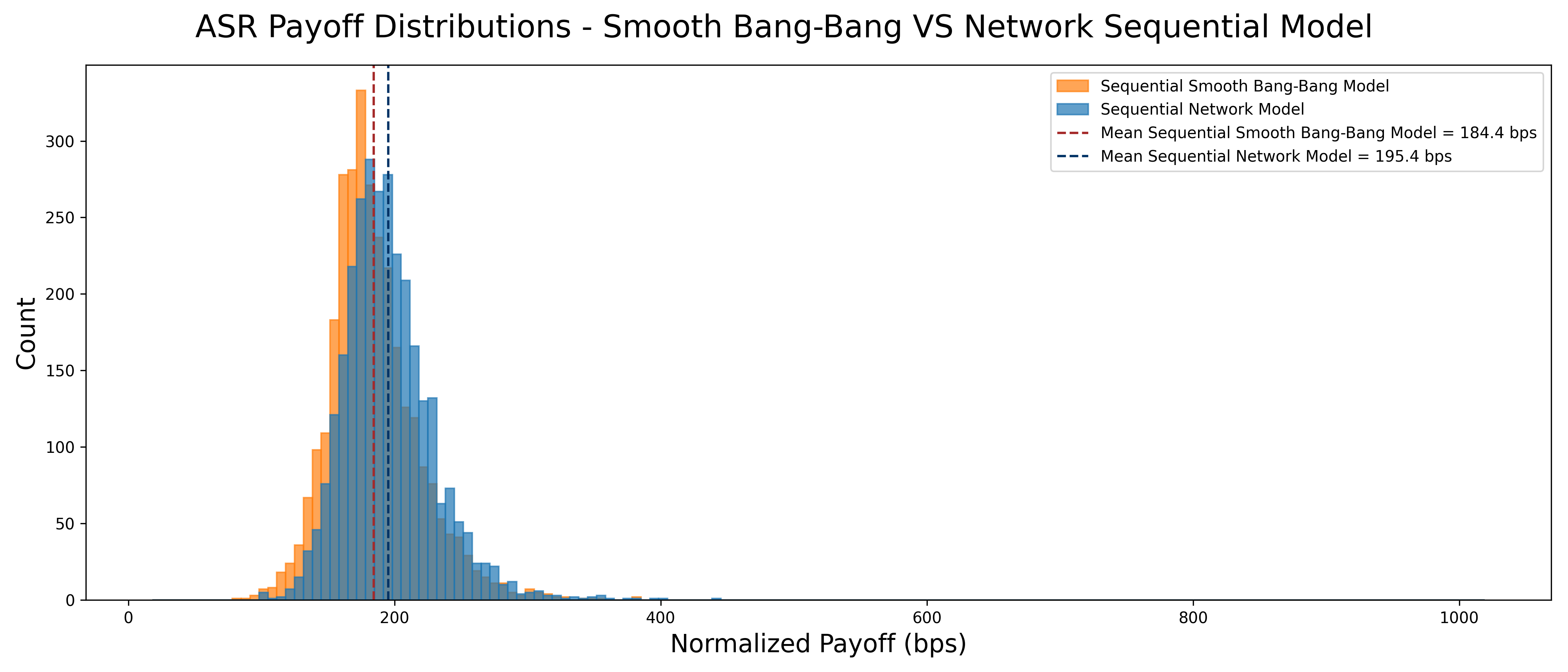} 
    \caption{Payoff distribution for the \textit{sequential model} applied to the \textit{smooth bang-bang} and \textit{network} policies. The payoff $PnL = PnL^{ASR} + PnL^{Hedge}$ is normalized by $W_{Min}$ and expressed in basis points.} 
    \label{fig:sequentialFig} 
\end{figure}

As is evident from the figure, the two \textit{sequential} models exhibit similar distributions, even if for the \textit{smooth bang-bang} it is slightly more peaked, indicating that the hedging framework achieves comparable final results. This is further supported by the similar values of $ES_{0.75}$, $MV_{2.5 \cdot 10^2}$, and $\delta^{ind}$. The advantage of the \textit{network} policy can be attributed to its superior performance in managing the contract, as reflected in the higher $\delta^{fair}$.

\vspace{0.1 cm}
We now analyze the results obtained from the \textit{joint} hedging models, showing the distribution of $PnL^{ASR}_\tau$, excluding $PnL^{Hedge}_\tau$, as it provides crucial insights. We compare the histogram of the ASR payoff obtained without hedging the repurchase program with the one obtained under \textit{joint} hedging. In the former case, the agent does not assume that hedging will occur and therefore must search for a trade-off between profit and risk. Specifically, the resulting distribution is the one presented in Section \ref{sec:buyback}, identical to the \textit{sequential} approach in which the repurchase implementation has been fixed. This comparison allows us to assess whether the possibility of taking risks that will ultimately be hedged away enables the bank to enhance profitability.

\begin{figure}[H] 
    \centering
    \includegraphics[width=0.9\textwidth]{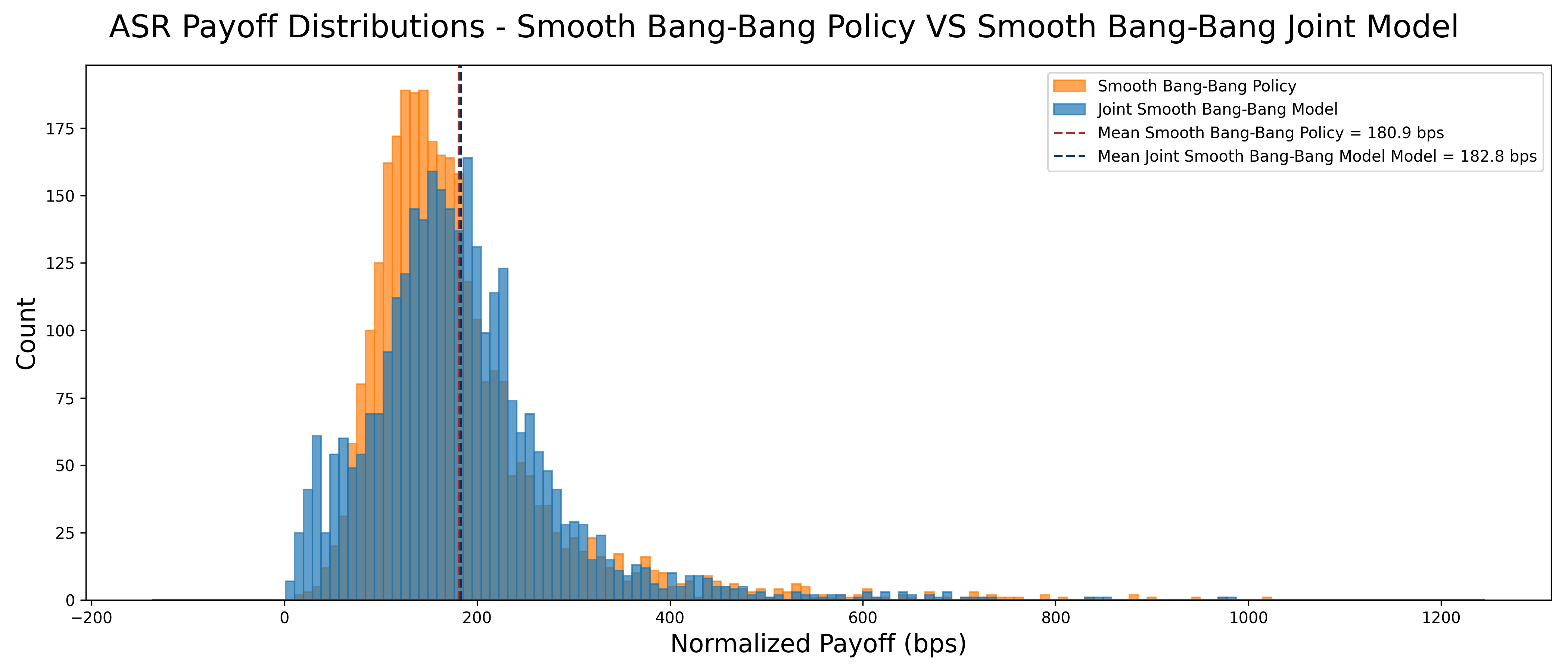} 
    \caption{Distribution of $PnL^{ASR}$: in orange it is represented the \textit{smooth bang-bang} in absence of the hedging portfolio, as in Section \ref{sec:buyback}. In light-blue is represented the \textit{joint} hedging model. The payoff is normalized by $W_{Min}$ and expressed in basis points.} 
    \label{fig:jointSmooth} 
\end{figure}

\vspace{0.1 cm}
Figure \ref{fig:jointSmooth} refers to the \textit{smooth bang-bang} policy. The light-blue distribution, obtained when the bank jointly manages the repurchase and its associated risks, displays a heavier left tail, which can be offset through the portfolio that generates $PnL^{Hedge}_\tau$. At the same time, the overall distribution becomes less peaked and the right tail is more pronounced. In other words, the joint framework allows for a more flexible and permissive approach to ASR execution. Nevertheless, the values of $PnL^{ASR}_\tau$ remain strictly positive, and the increase in the mean value is modest, rising from $180.9$ bps to $182.8$ bps.

\begin{figure}[H] 
    \centering
    \includegraphics[width=0.9\textwidth]{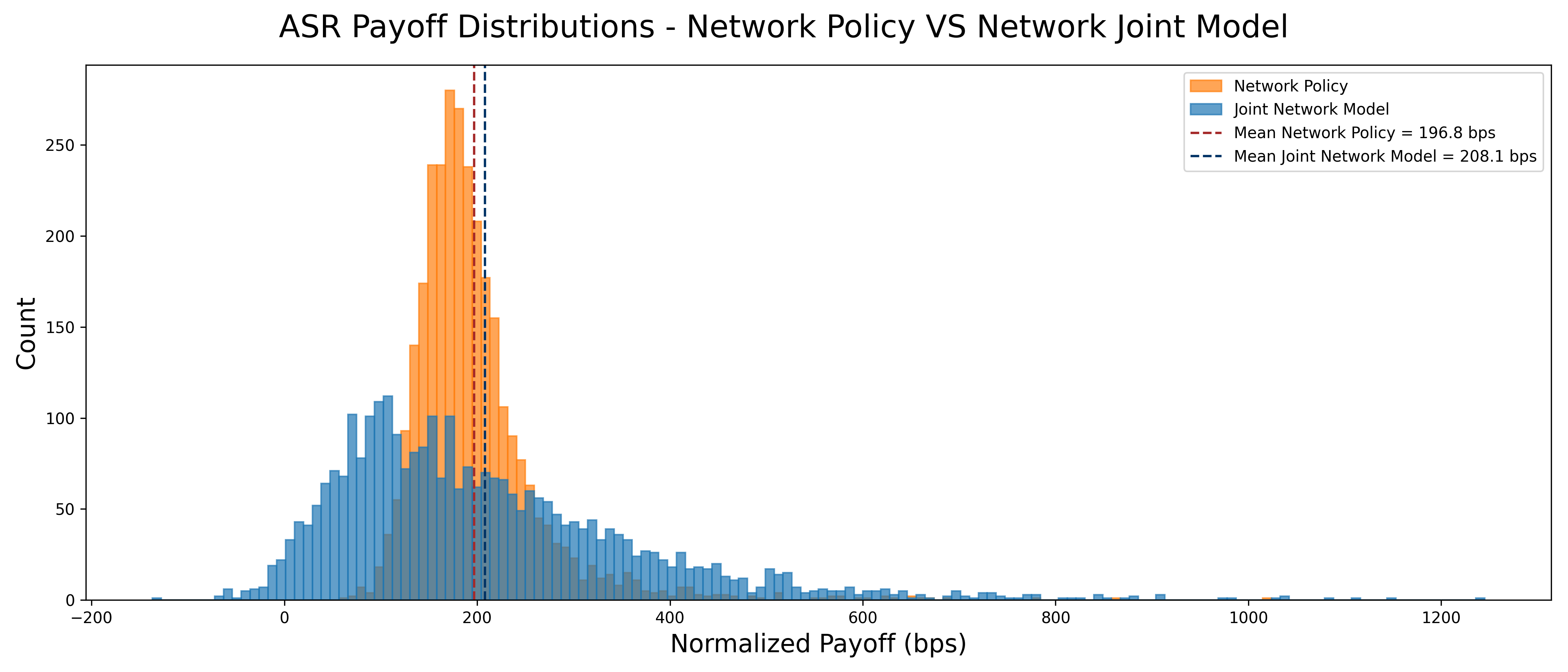} 
    \caption{Distribution of $PnL^{ASR}$: in orange it is represented the \textit{network} policy in absence of the hedging portfolio, as in Section \ref{sec:buyback}. In light-blue is represented the \textit{joint} hedging model. The payoff is normalized by $W_{Min}$ and expressed in basis points.} 
    \label{fig:networkJoint} 
\end{figure}

Figure \ref{fig:networkJoint}, related to the \textit{network} policy, shows similar effects, but on a larger scale. The distribution becomes significantly flatter, and the left tail now includes negative values. This can be interpreted as a consequence of the greater flexibility provided by the policy itself, which enables the agent to execute repurchases much more aggressively. This is also reflected in the increase in the mean value, which rises from $196.8$ bps to $208.1$ bps.

The tables below summarize the models considered so far, allowing us to assess the effectiveness of the overall framework. As shown in the first two tables, corresponding to the expected shortfall and mean-variance optimizations, respectively, there are two clear directions along which the risk measures decrease.

\vspace{0.1 cm}
The first is the horizontal direction: moving from left to right, the \textit{network} policy consistently outperforms the corresponding model implemented with the \textit{smooth bang-bang}. This relationship is partly expected, as the former is inherently more complex and relies on a substantially larger set of parameters. However, such added complexity must be justified by tangible improvements, which in practice can be evaluated through the discounts offered to the client. 

\vspace{0.1 cm}
The second direction of improvement is vertical: moving from top to bottom highlights the advantages of incorporating hedging either on its own or, more effectively, within a unified framework. Nevertheless, the magnitude of the reduction from the \textit{sequential} to the \textit{joint} hedging approach depends strongly on the underlying policy, as anticipated by Figures \ref{fig:jointSmooth} and \ref{fig:networkJoint}. The \textit{smooth bang-bang} strategy exhibits only a modest decrease, indicating that handling the repurchase and the hedging components jointly does not significantly affect its performance. In contrast, the network policy benefits considerably from an integrated workflow; its greater flexibility leads to a pronounced improvement when both tasks are optimized simultaneously.
\begin{table}[H]
\centering
\begin{tabular}{|c|c|c|c|}
\hline
$ES_{0.75}$ & \textit{Smooth Bang-Bang} & \textit{Network} policy \\
\hline
No Hedging         & $-94.0$  & $-136.8$ \\
\hline
Sequential Hedging & $-147.4$ & $-158.1$ \\
\hline
Joint Hedging      & $-148.3$ & $-171.4$ \\
\hline
\end{tabular}
\caption{Expected shortfall for the all hedging models with the \textit{smooth bang-bang} and the \textit{network} policies, expressed in basis points. The \textit{No Hedging} values are those presented in Section \ref{sec:buyback}.}
\label{tab:finalES}
\end{table}

\begin{table}[H]
\centering
\begin{tabular}{|c|c|c|c|}
\hline
$MV_{2.5 \cdot 10^2}$ & \textit{Smooth Bang-Bang} & \textit{Network} policy \\
\hline
No Hedging         & $-62.8$  & $-161.1$ \\
\hline
Sequential Hedging & $-162.2$ & $-174.5$ \\
\hline
Joint Hedging      & $-171.6$ & $-195.7$ \\
\hline
\end{tabular}
\caption{Mean-variance for the all the hedging models with the \textit{smooth bang-bang} and the \textit{network} policies, expressed in basis points. The \textit{No Hedging} values are those presented in Section \ref{sec:buyback}.}
\label{tab:finalMV}
\end{table}

We now turn to the pricing system based on discounts on the benchmark price. The \textit{joint} hedging model increases the fair discount $\delta^{fair}$ for both ASR strategies compared to the sequential approach. As discussed above, this improvement comes from the model’s ability to mitigate risk: by hedging effectively, the agent can afford to take on additional exposure while repurchasing shares in the market. This greater freedom allows for a more aggressive execution, ultimately leading to higher profitability. 
\begin{table}[H]
\centering
\begin{tabular}{|c|c|c|c|}
\hline
$\delta^{fair}$ & \textit{Smooth Bang-Bang} & \textit{Network} policy \\
\hline
No Hedging         & $156.5$ & $167.4$ \\
\hline
Sequential Hedging & $156.5$ & $167.4$ \\
\hline
Joint Hedging      & $160.8$ & $175.3$ \\
\hline
\end{tabular}
\caption{Fair discount for the all the hedging models with the \textit{smooth bang-bang} and the \textit{network} policies, expressed in basis points. The \textit{No Hedging} values are those presented in Section \ref{sec:buyback}.}
\label{tab:finalDeltaFair}
\end{table}

The indifference discount $\delta^{ind}$ follows the same pattern: even when the product is priced incorporating the actual risk faced by the bank through the risk measure $\rho$, the \textit{joint} models achieve superior performance, allowing the bank to offer a higher sustainable discount.

\vspace{0.1 cm}
Further insights emerge when comparing the two ASR strategies, third column of Table \ref{tab:finalDeltaInd}. Although the sequential model narrows the difference in indifference discounts, this gap widens once the optimization becomes unified. This confirms our initial intuition: a more flexible policy benefits to a greater extent from a unified optimization process, as the potential to establish a dependence between the two components of the final payoff can be fully realized only if the agent is able to adapt its execution. However, this requires the ability to incorporate and exploit a richer representation of the state space.

\begin{table}[H]
\centering
\begin{tabular}{|c|c|c|c|c|}
\hline
$\delta^{ind}$ & \textit{Smooth Bang-Bang} & \textit{Network} policy & Difference\\
\hline
No Hedging         & $82.1$  & $114.6$ & $32.5$\\
\hline
Sequential Hedging & $126.5$ & $132.4$ & $5.9$\\
\hline
Joint Hedging      & $127.1$ & $142.3$ & $15.2$\\
\hline
\end{tabular}
\caption{Indifference discount for the all the hedging models with the \textit{smooth bang-bang} and the \textit{network} policies, expressed in basis points.  The \textit{No Hedging} values are those presented in Section \ref{sec:buyback}.}
\label{tab:finalDeltaInd}
\end{table}

We conclude by examining the gap between the optimal theoretical hedging performance and the realized one, measured by $\delta^{fair} - \delta^{ind}$. As previously discussed, the \textit{sequential} hedging model reduces this gap, yielding comparable values for the two ASR strategies. However, the \textit{joint} model does not significantly alter the situation, leaving $\delta^{fair} - \delta^{ind}$ essentially unchanged. In our view, this is due to the fact that the hedging network architecture is the same across models and therefore has the same computational capacity, limiting the achievable improvement in this dimension.
\begin{table}[H]
\centering
\begin{tabular}{|c|c|c|c|}
\hline
$\delta^{fair} - \delta^{ind}$ & \textit{Smooth Bang-Bang} & \textit{Network} policy \\
\hline
No Hedging         & $74.4$  & $52.8$ \\
\hline
Sequential Hedging & $30.0$ & $35.0$ \\
\hline
Joint Hedging      & $33.7$ & $33.0$ \\
\hline
\end{tabular}
\caption{Difference in discounts for the all the hedging models with the \textit{smooth bang-bang} and the \textit{network} policies, expressed in basis points.  The \textit{No Hedging} values are those presented in Section \ref{sec:buyback}.}
\label{tab:finalDeltaDif}
\end{table}

In conclusion, we can answer positively one of the key questions underlying this work: a unified framework is indeed advantageous for the financial institution, and this advantage becomes increasingly evident as the implementation becomes more flexible. The two plots below show the overall improvement by representing $PnL^{ASR}_\tau$ for the policies presented in Section 3 and $PnL_\tau = PnL^{ASR}_\tau + PnL^{Hedge}_\tau$ for the \textit{joint} models. For the \textit{smooth bang-bang}: 
\begin{figure}[H] 
    \centering
    \includegraphics[width=0.9\textwidth]{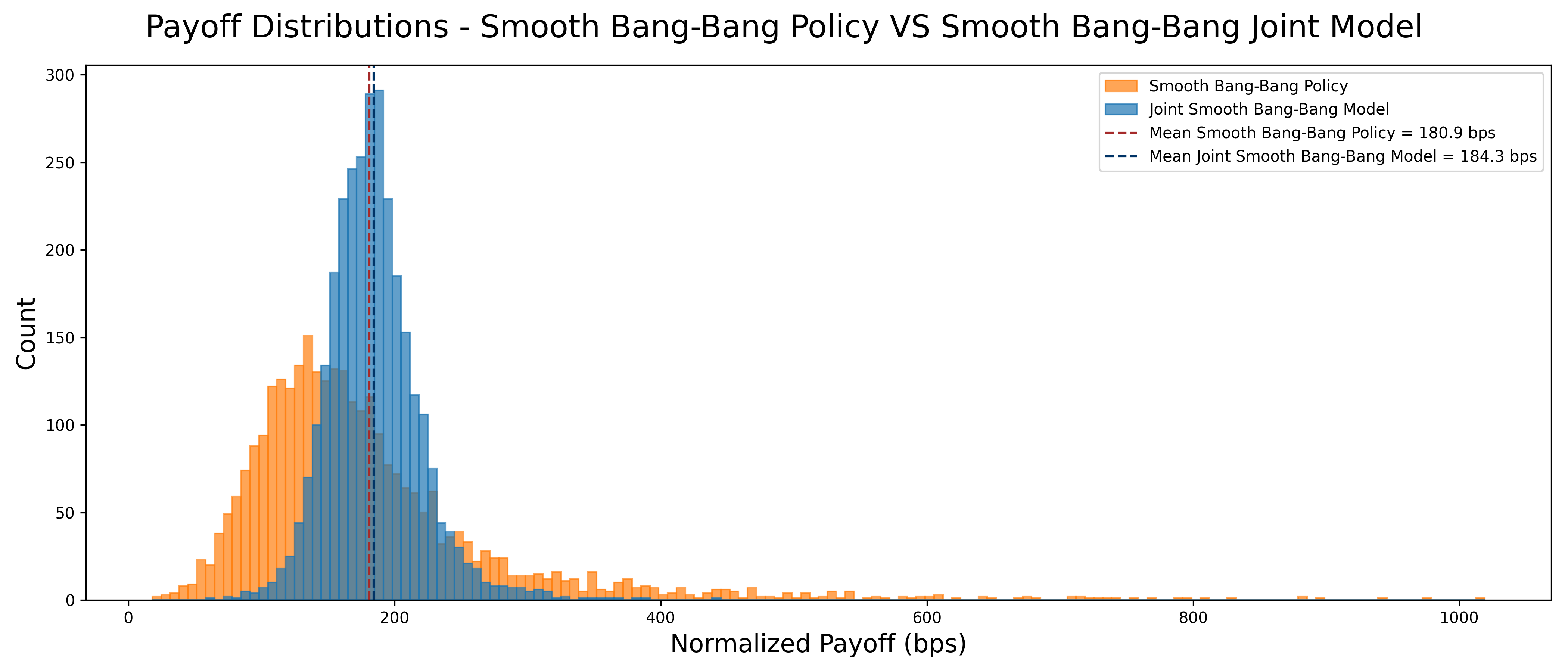} 
    \caption{Distribution of $PnL = PnL^{ASR} + PnL^{Hedge}$: in orange it is represented the \textit{smooth bang-bang} in absence of the hedging portfolio, as in Section \ref{sec:buyback}. In light-blue is represented the \textit{joint} hedging model. The payoff is normalized by $W_{Min}$ and expressed in basis points.} 
\end{figure}
\noindent
while the \textit{netowrk} one yields:
\begin{figure}[H] 
    \centering
    \includegraphics[width=0.9\textwidth]{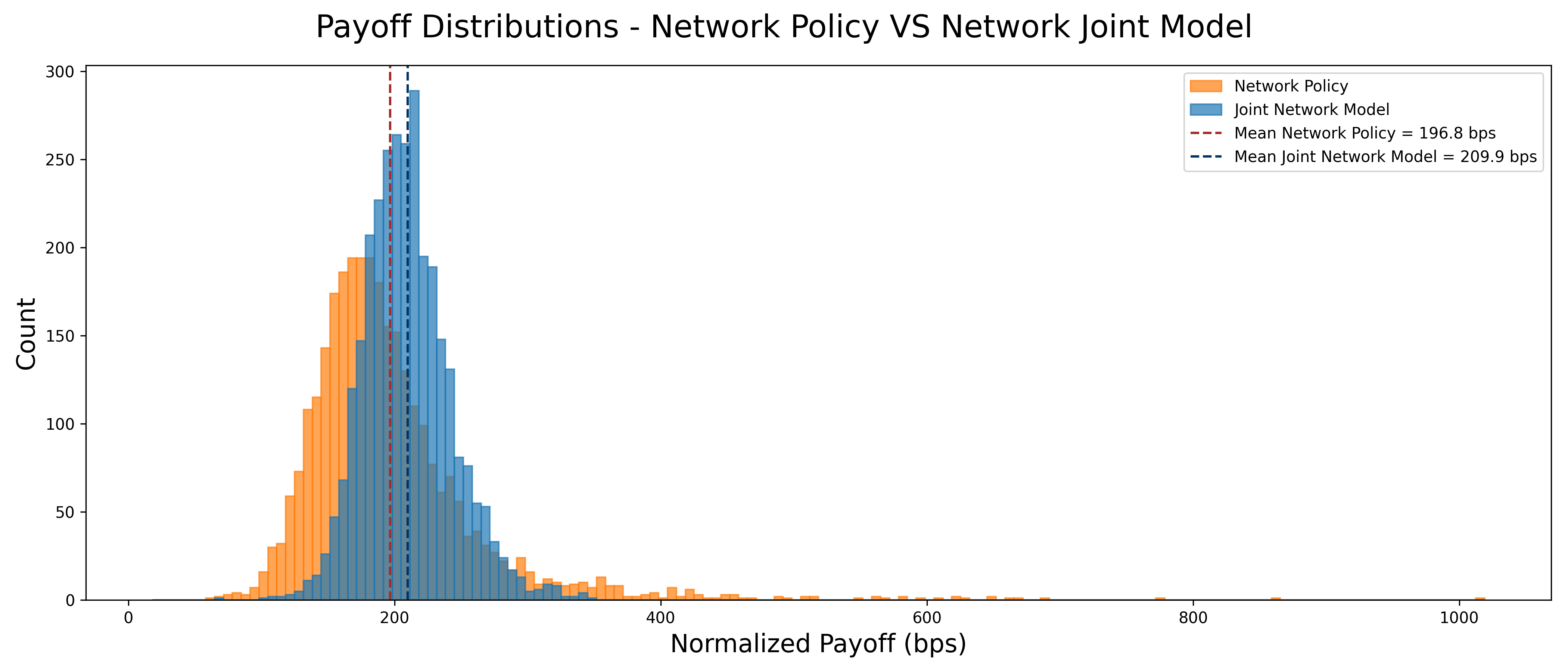} 
    \caption{Distribution of $PnL = PnL^{ASR} + PnL^{Hedge}$: in orange it is represented the \textit{network} in absence of the hedging portfolio, as in Section \ref{sec:buyback}. In light-blue is represented the \textit{joint} hedging model. The payoff is normalized by $W_{Min}$ and expressed in basis points.} 
    \label{fig:finalNet} 
\end{figure}

\subsection{Hedging Constraints}
In the final stage of our analysis, we develop a model that incorporates the full set of constraints \ref{eq:constraints} described in Section 2.2. Since the second limitation determines a dependence between the repurchase activity and its associated hedge, it is natural to build upon the \textit{joint} hedging framework, as the policy can adapt optimally to the contract’s clauses.

\vspace{0.1 cm}
Unlike in previous models, the number of shares $h_n$ held at time $t_n$ can no longer take values in $\mathbb{R}$. Instead, it is restricted to the interval:
$$
h_n \in [h_{n-1} - (\overline{\nu}_n^h - b_n), \text{ } h_{n-1} + \overline{\nu}_n^h - b_n] = \mathcal{I}_n^h \quad \forall n \in \mathcal{T} \setminus \{ N_{Max} \}
.$$

\vspace{0.1 cm}
This condition is achieved by capping the output of the hedging network $h_n$ to $\mathcal{I}_n^h$:
$$
h_n = \min\left(h_{n-1} + \overline{\nu}_n^h - b_n,\text{ } \max\left(h_{n-1} - (\overline{\nu}_n^h - b_n), \text{ } h_n^*\right)\right)
,$$
$$
\text{where} \quad h_n^* = \lambda \cdot f_{\theta}\left(\frac{n - N_{Min}}{N_{Min}}, \text{ } \frac{S_n}{A_n}, \text{ } \frac{W_{n-1}}{W_{Min}}\right) \quad \text{and} \qquad \lambda = \frac{W_{Min} + W_{Max}}{S_0 \cdot (N_{Min} + N_{Max})}.
$$

\vspace{0.1 cm}
The architecture of $f_\theta$ is the same of the previous section. As an alternative, one may use a bounded activation on the final layer, for example the sigmoid, and map the output to $\mathcal{I}_n^h$. However, our experiments suggest that this change in the architecture of $f_\theta$ leads to suboptimal outcomes.

\vspace{0.1 cm}
We now present the numerical results. Similarly to the processes $\underline{\nu}_n \text{ and } \overline{\nu}_n$, we consider constant values $\overline{\nu}_n^h$ over time, and we examine the scenarios corresponding to $\overline{\nu}^h \in \{1.5 \cdot 10^6, \text{ } 2 \cdot 10^6, \text{ } 2.5 \cdot 10^6 \}$. In the first scenario, $\overline{\nu}^h = \overline{\nu}$, resulting in identical upper bounds for $b_n$ and for the sum $b_n + |h_n - h_{n-1}|$.

\vspace{0.1 cm}
We consider the following three models, whose repurchase activity is delegated to the \textit{network} policy:
\begin{itemize}[noitemsep, topsep=0pt]
    \item the \textit{free} hedging model, corresponding to the \textit{joint} hedging formulation introduced earlier, in which $h_n \in \mathbb{R}$
    \item the \textit{capped} hedging model, where the value of $h_n$ is obtained by capping the output of the \textit{free} model to the set $\mathcal{I}_n^h$. It is used to assess how an ex post implementation performs
    \item the \textit{constrained} hedging model, in which the optimization is performed under the constraint $h_n \in \mathcal{I}_n^h$. The difference with respect to the \textit{capped} model is that the capping mechanism is now applied at optimization time, allowing the agent to learn to behave optimally under the real trading capabilities.
\end{itemize}

\vspace{0.1 cm}
We first compare the risk measures obtained by the \textit{constrained} implementation. As shown in Table \ref{tab:riskMeasuresConstrained}, the agent experiences worse outcomes in terms of risk when the additional constraints are imposed. The row $\overline{\nu}^h = + \infty$ corresponds to the \textit{free} model.
\begin{table}[H]
\centering
\begin{tabular}{|c|c|c|c|c|c|c|}
\hline
$\overline{\nu}^h$ & $ES_{0.75}$ & $MV_{2.5 \cdot 10^2}$ \\
\hline
$1.5 \cdot 10^6$   & $-163.3$ & $-189.7$ \\
\hline
$2 \cdot 10^6$   & $-167.3$ & $-192.6$ \\
\hline
$2.5 \cdot 10^6$ & $-169.4$ & $-194.4$ \\
\hline
$+ \infty$    & $-171.4$ & $-195.7$ \\
\hline
\end{tabular}
\caption{Risk measures for the \textit{constrained} hedging models, expressed in basis points. The \textit{free} corresponds to $\overline{\nu}^h = + \infty$.}
\label{tab:riskMeasuresConstrained}
\end{table}

Although the increased complexity of the intermediary’s task may directly contribute to the reduction in performance, additional factors may also play a role. We thus consider the expected shortfall and the discounts offered on the benchmark price for each value of $\overline{\nu}^h$. We note that the value of $\delta^{fair}$ is identical for the \textit{free} and \textit{capped} models, as they share the same repurchase strategy parameters $\phi$. 

\begin{table}[H]
\centering
\begin{tabular}{|c|c|c|c|c|c|}
\hline
$\overline{\nu}^h = 1.5 \cdot 10^6$ & $ES_{0.75}$ & $\delta^{fair}$ & $\delta^{ind}$ & $\delta^{fair} - \delta^{ind}$ \\
\hline
\textit{Free}    & $-171.4$ & $175.3$ & $142.3$ & $33.0$\\
\hline
\textit{Capped}   & $-127.7$ & $175.3$ & $104.5$ & $70.8$ \\
\hline
\textit{Constrained}   & $-163.3$ & $172.7$ & $135.8$ & $36.9$ \\
\hline
\end{tabular}
\caption{Discount for \textit{free}, \textit{capped} and \textit{constrained} hedging models for $\overline{\nu}^h = 1.5 \cdot 10^6$, in bps}
\label{tab:ConstOnePointFiveDisc}
\end{table}

\begin{table}[H]
\centering
\begin{tabular}{|c|c|c|c|c|c|}
\hline
$\overline{\nu}^h = 2 \cdot 10^6 $ & $ES_{0.75}$ & $\delta^{fair}$ & $\delta^{ind}$ & $\delta^{fair} - \delta^{ind}$ \\
\hline
\textit{Free}    & $-171.4$ & $175.3$ & $142.3$ & $33.0$\\
\hline
\textit{Capped}   & $-163.7$ & $175.3$ & $135.0$ & $40.3$ \\
\hline
\textit{Constrained}   & $-167.3$ & $174.4$ & $139.1$ & $35.3$ \\
\hline
\end{tabular}
\caption{Discount for \textit{free}, \textit{capped} and \textit{constrained} hedging models for $\overline{\nu}^h = 2 \cdot 10^6$, in bps.}
\label{tab:ConstTwoDisc}
\end{table}

\begin{table}[H]
\centering
\begin{tabular}{|c|c|c|c|c|c|}
\hline
$\overline{\nu}^h = 2.5 \cdot 10^6 $ & $ES_{0.75}$ & $\delta^{fair}$ & $\delta^{ind}$ & $\delta^{fair} - \delta^{ind}$ \\
\hline
\textit{Free}    & $-171.4$ & $175.3$ & $142.3$ & $33.0$\\
\hline
\textit{Capped}   & $-168.8$ & $175.3$ & $139.7$ & $35.6$ \\
\hline
\textit{Constrained}   & $-169.4$ & $174.9$ & $140.7$ & $34.2$ \\
\hline
\end{tabular}
\caption{Discount for \textit{free}, \textit{capped} and \textit{constrained} hedging models for $\overline{\nu}^h = 2.5 \cdot 10^6$, in bps.}
\label{tab:ConstTwoPointFiveDisc}
\end{table}

In general, the \textit{constrained} model outperforms the \textit{capped} in terms of both expected shortfall and indifference discount for all values of $\overline{\nu}^h$. This shows that the agent can extract more value from the buyback if its strategy is optimized while considering trading constraints, rather than imposing them ex post.

\vspace{0.1 cm}
Specifically, the improvement in performance of the \textit{constrained} model becomes increasingly evident as the value of $\overline{\nu}^h$ decreases. The poor results of the \textit{capped} implementation reported in Table \ref{tab:ConstOnePointFiveDisc} highlight the inadequacy of enforcing the constraints only ex post. The indifference discount $\delta^{ind}$ drops dramatically compared to $\delta^{fair}$, showing that the latter is a poor performance estimator when the agent operates under restricted capabilities. By contrast, as $\overline{\nu}^h$ increases, the performance gap decreases. 

\begin{figure}[H] 
    \centering
    \includegraphics[width=0.9\textwidth]{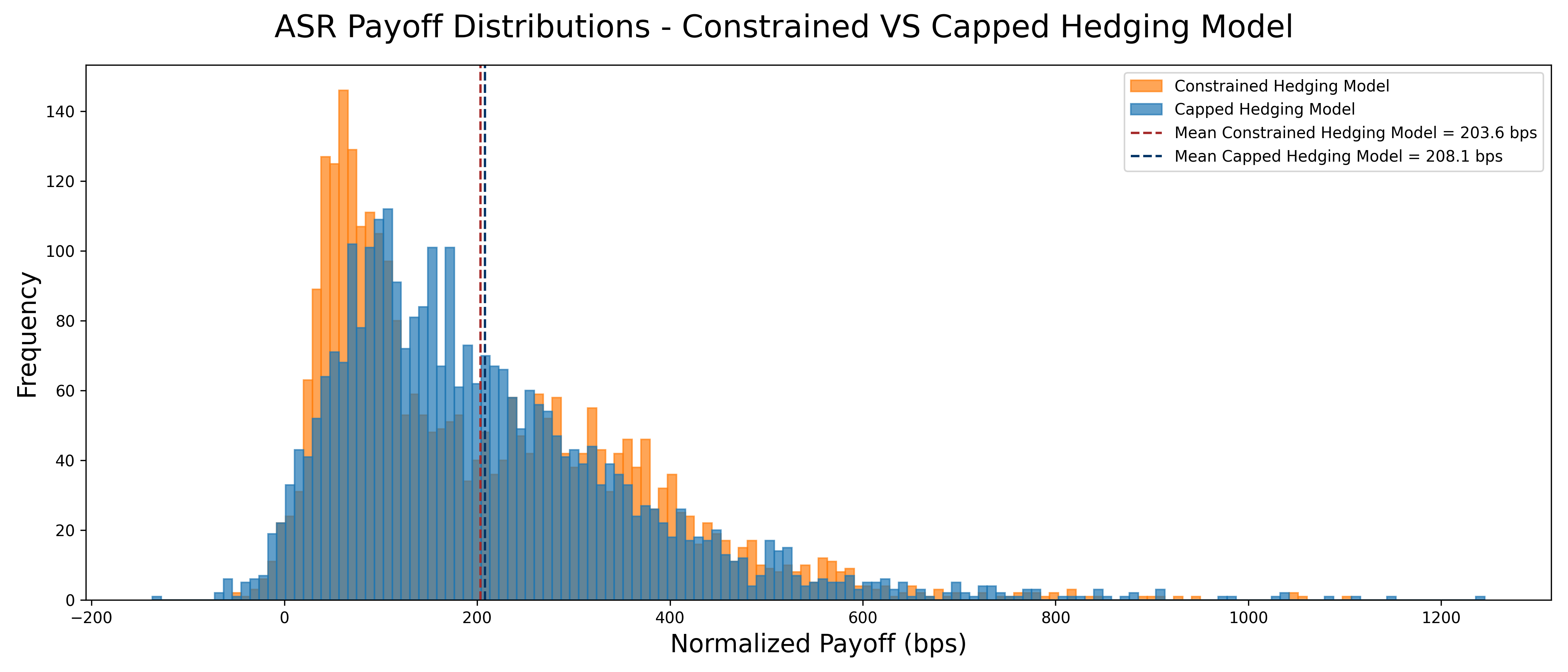} 
    \caption{Distribution of $PnL^{ASR}$: in orange it is represented the \textit{constrained} model, while in light-blue the \textit{capped}. The payoff is normalized by $W_{Min}$ and expressed in basis points.} 
    \label{fig:constrainedASR} 
\end{figure}

\vspace{0.1 cm}
We present the distribution plots comparing the \textit{constrained} and \textit{capped} hedging models for $\overline{\nu}^h = 1.5 \cdot 10^6$, value for which the two approaches differ the most. However, as $\overline{\nu}^h$ increases, the two implementation show similar distribution plots. Figure \ref{fig:constrainedASR} reports the distribution of $PnL^{ASR}_\tau$. The \textit{constrained} distribution, shown in orange, is bimodal. The right tale is heavier than the \textit{capped} and, on the left tail, few are the paths in which the repurchase leads to a negative payoff. This may be a consequence of accounting for the actual limitations on the hedging activity at optimization time, which possibly pushes to the adoption of a more conservative approach. 

\begin{figure}[H] 
    \centering
    \includegraphics[width=0.9\textwidth]{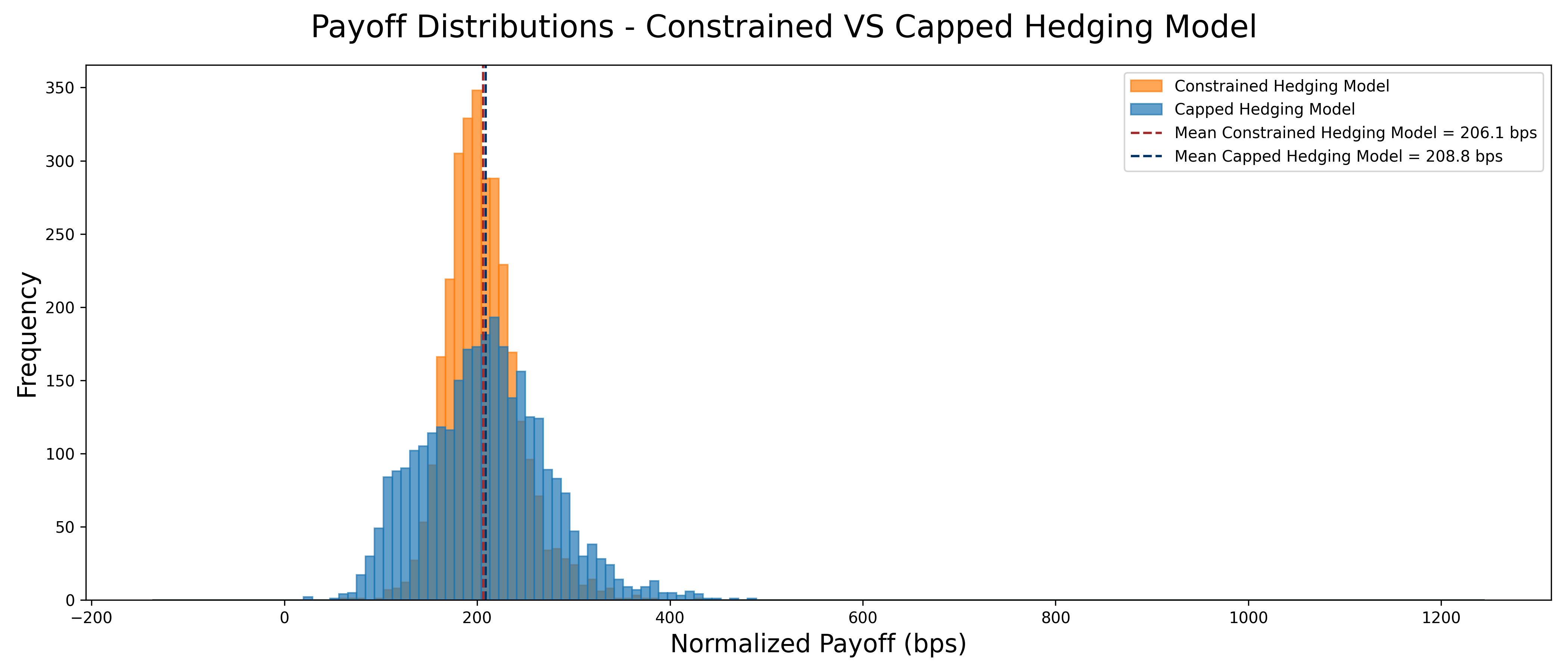} 
    \caption{Distribution of $PnL = PnL^{ASR} + PnL^{Hedge}$: in orange it is represented the \textit{constrained} model, while in light-blue the \textit{capped}. The payoff is normalized by $W_{Min}$ and expressed in basis points.} 
    \label{fig:constrainedFull} 
\end{figure}

\vspace{0.1 cm}
Figure \ref{fig:constrainedFull} displays the distribution of the total profit and loss, $PnL_\tau = PnL^{ASR}_\tau + PnL^{Hedge}_\tau$. When the constraint $h_n \in \mathcal{I}^h$ is imposed ex post, the \textit{capped} model fails to fully eliminate the exposure, as evidenced by its heavier tails.

In conclusion, explicitly accounting for hedging constraints increases the complexity of the optimal control problem but provides a meaningful advantage, which is more evident when the agent has less trading capabilities.

\section{Conclusion}
In this work, we address the management of repurchase programs and their corresponding hedging through machine learning techniques, with a particular focus on neural networks. We examine a standard contract that incorporates features commonly found in the buyback market, such as the greenshoe on the notional amount, expressed in cash, and trading constraints.

\vspace{0.1 cm}
The flexibility granted to the buyback executor allows us to capture all dependencies between the two activities, showing that a unified framework for both tasks yields substantial performance improvements. This can be effectively leveraged to address trading constraints that simultaneously affect both activities, providing a more realistic hedging approach, in the sense that the bank’s policy is optimized while accounting for actual trading capabilities.

\vspace{0.1 cm}
Since policy evaluation is delegated to risk measures, as is standard in the literature, we rely on an alternative and more realistic indifference pricing approach that reflects the trading capacity. The results confirm the effectiveness of a joint optimization framework for simultaneously managing the product and its hedge, also when trading limitations are considered.

\vspace{0.1 cm}
Several directions for future research arise from our analysis. The framework can be extended to incorporate exotic contractual features, such as cap and floor structures or performance-sharing mechanisms, which are often present in ASR agreements and pose challenges for standard techniques such as Bellman-based dynamic programming. Another promising avenue is the adoption of more realistic stochastic models, particularly those that include stochastic drivers for volatility, to better capture market dynamics and improve the robustness of the resulting policies. 

\section*{Disclaimer}

The authors report no potential competing interests. The opinions expressed in this document are solely those of the authors and do not represent in any way those of their present and past employers.

\printbibliography

@article{Miller3,
    author =       "Miller, M. and Modigliani, F.",
    title =        "Dividend Policy, Growth, and the Valuation of Shares",
    journal =      "The Journal of Business",
    year =         "1961",
    volume =       "34", 
    pages =        "411--433",
    DOI =          "10.1086/294442"
}

@incollection{Allen4,
    author =       "Allen, F. and Michaely, R.",
    title =        "Payout policy",
    booktitle = {Handbook of the Economics of Finance},
    editor    = {Constantinides, G. M. and Harris, M. and Stulz, R. M.},
    publisher = {Elsevier},
    year      = {2003},
    pages     = {337--429},
    volume    = {1},
    edition   = {1},
    chapter   = {7}
}

@article{Farre-Mensa5,
  author  = {Farre-Mensa, J. and Michaely, R. and Schmalz, M.},
  title   = {Payout Policy},
  journal = {Annual Review of Financial Economics},
  volume  = {6},
  number  = {1},
  pages   = {75--134},
  year    = {2014},
  doi     = {10.1146/annurev-financial-110613-034259}
}

@article{Gueant6,
  author  = {Guéant, O. and Manziuk, I. and Pu, J.},
  title   = {Accelerated share repurchase and other buyback programs: what neural networks can bring},
  journal = {Quantitative Finance},
  volume  = {20},
  number  = {8},
  pages   = {1389--1404},
  year    = {2020},
  doi     = {10.1080/14697688.2020.1729397}
}

@article{Buehler7,
  author  = {Buehler, H. and Gonon, L. and Teichmann, J. and Wood, B.},
  title   = {Deep hedging},
  journal = {Quantitative Finance},
  volume  = {19},
  number  = {8},
  pages   = {1271--1291},
  year    = {2019},
  doi     = {10.1080/14697688.2019.1571683}
}

@article{Mikkila8,
  author  = {Mikkilä, O. and Kanniainen, J.},
  title   = {Empirical deep hedging},
  journal = {Quantitative Finance},
  volume  = {23},
  number  = {1},
  pages   = {111--122},
  year    = {2023},
  doi     = {10.1080/14697688.2022.2136037},
  url     = {https://trepo.tuni.fi/handle/10024/222907}
}

@article{Baldacci10,
  author  = {Baldacci, B. and Bergault, P. and Gu{\'e}ant, O.},
  title   = {Pricing and Managing Complex Share Buy-Back Contracts: An Alternative to Optimal Control},
  journal = {Risk},
  year    = {2024},
  date    = {2024-08-14},
  url     = {https://www.risk.net/cutting-edge/7959784/pricing-and-managing-complex-share-buy-back-contracts-an-alternative-to-optimal-control},
  note    = {Published online}
}

@book{Foellmer11,
  author    = {F{\"o}llmer, H. and Schied, A.},
  title     = {Stochastic Finance: An Introduction in Discrete Time},
  publisher = {De Gruyter},
  year      = {2016},
  edition   = {4},
  doi       = {10.1515/9783110463453},
  isbn      = {9783110463453}
}

@article{Gueant12,
  author  = {Gu{\'e}ant, O. and Pu, J. and Royer, G.},
  title   = {Accelerated Share Repurchase: pricing and execution strategy},
  journal = {International Journal of Theoretical and Applied Finance},
  volume  = {18},
  number  = {3},
  pages   = {1--31},
  year    = {2015},
  doi     = {10.1142/S0219024915500193}
}

@article{Gueant13,
  author  = {Gu{\'e}ant, Olivier},
  title   = {Optimal Execution of Accelerated Share Repurchase Contracts with Fixed Notional},
  journal = {Journal of Risk},
  volume  = {19},
  number  = {5},
  pages   = {77--99},
  year    = {2017},
  doi     = {10.21314/JOR.2017.361}
}

@article{Foellmer14,
  author  = {F{\"o}llmer, H. and Schied, A.},
  title   = {Convex Measures of Risk and Trading Constraints},
  journal = {Finance and Stochastics},
  volume  = {6},
  pages   = {429--447},
  year    = {2002},
  month   = {09},
  doi     = {10.1007/s007800200072}
}

@article{BenTal15,
  author  = {Ben-Tal, A. and Teboulle, M.},
  title   = {An old-new concept of convex risk measures: the optimized certainty equivalent},
  journal = {Mathematical Finance},
  volume  = {17},
  number  = {3},
  pages   = {449--476},
  year    = {2007},
  doi     = {10.1111/j.1467-9965.2007.00311.x}
}

@misc{Pickard16,
  author       = {Pickard, R. and Wredenhagen, F. and DeJesus, J. and Schlener, M. and Lawryshyn, Y.},
  title        = {Hedging American Put Options with Deep Reinforcement Learning},
  year         = {2024},
  eprint       = {2405.06774},
  archivePrefix= {arXiv},
  primaryClass = {q-fin.CP},
  note         = {arXiv preprint},
  url          = {https://arxiv.org/abs/2405.06774}
}

@misc{Oya17,
  author       = {Oya, K.},
  title        = {Deep Hedging Bermudan Swaptions},
  year         = {2024},
  eprint       = {2411.10079},
  archivePrefix= {arXiv},
  primaryClass = {q-fin.CP},
  note         = {arXiv preprint},
  url          = {https://arxiv.org/abs/2411.10079}
}

@article{Horikawa22,
  author  = {Horikawa, H. and Nakagawa, K.},
  title   = {Relationship between deep hedging and delta hedging: Leveraging a statistical arbitrage strategy},
  journal = {Finance Research Letters},
  volume  = {62},
  pages   = {105101},
  year    = {2024},
  issn    = {1544-6123},
  doi     = {10.1016/j.frl.2024.105101}
}

@misc{Francois23,
  author       = {Fran\c{c}ois, P. and Gauthier, G. and Godin, F. and P\'erez-Mendoza, C. O.},
  title        = {Deep Hedging with Options Using the Implied Volatility Surface},
  year         = {2025},
  eprint       = {2504.06208},
  archivePrefix= {arXiv},
  primaryClass = {q-fin.RM},
  note         = {arXiv preprint},
  url          = {https://arxiv.org/abs/2504.06208}
}

@misc{Armstrong24,
  author       = {Armstrong, J. and Tatlow, G.},
  title        = {Deep Gamma Hedging},
  year         = {2024},
  eprint       = {2409.13567},
  archivePrefix= {arXiv},
  note         = {arXiv preprint},
  url          = {https://arxiv.org/abs/2409.13567}
}

@article{Cao25,
  author  = {Cao, J. and Chen, J. and Farghadani, S. and Hull, J. and Poulos, Z. and Wang, Z. and Yuan, J.},
  title   = {Gamma and Vega Hedging using Deep Distributional Reinforcement Learning},
  journal = {Frontiers in Artificial Intelligence},
  volume  = {6},
  year    = {2023},
  doi     = {10.3389/frai.2023.1129370}
}

@article{Jaimungal32,
  author  = {Jaimungal, S. and Kinzebulatov, D. and Rubisov, D. H.},
  title   = {Optimal Accelerated Share Repurchases},
  journal = {Applied Mathematical Finance},
  year    = {2017},
  volume  = {24},
  number  = {3},
  pages   = {216--245},
  doi     = {10.1080/1350486X.2017.1374870}
}

\end{document}